\documentclass[fleqn,usenatbib]{mnras}

\usepackage{newtxtext,newtxmath}

\usepackage[T1]{fontenc}

\DeclareRobustCommand{\VAN}[3]{#2}
\let\VANthebibliography\thebibliography
\def\thebibliography{\DeclareRobustCommand{\VAN}[3]{##3}\VANthebibliography}

\usepackage{graphicx}	
\usepackage{amsmath}	
\usepackage{orcidlink}

\title[Rubin transient vs AGN selection]{Automatically distinguishing Rubin transients from AGN using variability metrics}

\author[D. Magill et al.]{
Dylan Magill$^{1}$\orcidlink{0009-0000-6521-8842}\thanks{E-mail: dmagill14@qub.ac.uk},
M. Nicholl$^{1}$\orcidlink{0000-0002-2555-3192},
P. Wiseman$^{2}$\orcidlink{0000-0002-3073-1512},
C. R. Angus$^{1}$\orcidlink{0000-0002-4269-7999},
P. Ramsden$^{3}$\orcidlink{0009-0009-2627-2884},
C. Frohmaier$^{4}$\orcidlink{0000-0001-9553-4723},
\newauthor
S. van Velzen$^{5}$\orcidlink{0000-0002-3859-8074},
V. Anilkumar$^{5}$\orcidlink{0009-0008-3146-287X},
J. Greenwood$^{6}$\orcidlink{0009-0008-7284-8234}
and J. G. Weston$^{1}$\orcidlink{0009-0002-9460-9900}
\\
$^{1}$Astrophysics Research Centre, School of Mathematics and Physics, Queen's University Belfast, Belfast BT7 1NN, UK\\
$^{2}$School of Physics and Astronomy, University of Southampton, Southampton SO17 1BJ, UK\\
$^{3}$School of Physics and Astronomy, University of Birmingham, Birmingham B15 2TT, UK\\
$^{4}$Institute of Cosmology and Gravitation, University of Portsmouth, Portsmouth, PO13FX, UK \\
$^{5}$Leiden Observatory, Leiden University, Postbus 9513, NL-2300 RA Leiden, the Netherlands \\
$^{6}$Jodrell Bank Centre for Astrophysics, School of Physics \& Astronomy, University of Manchester, Oxford Rd., Manchester M13 9PL, UK
}

\date{Accepted XXX. Received YYY; in original form ZZZ}

\pubyear{\the\year{}}

\begin{document}
\label{firstpage}
\pagerange{\pageref{firstpage}--\pageref{lastpage}}
\maketitle

\begin{abstract}

Stochastic variability of active galactic nuclei (AGN) can produce contaminants in the search for explosive extragalactic transients (such as supernovae and tidal disruption events). In the new era of the Rubin Observatory's Legacy Survey of Space and Time (LSST), previously uncatalogued AGN, especially those with luminosity near the survey detection limits, are expected to produce a flood of detections that have the potential to contaminate surveys targeting other transients, leading to inefficient use of spectroscopic follow-up time. For surveys aiming to statistically characterise transient demographics, it is advantageous to use easily modelled and reproducible selection criteria to distinguish AGN from other transients, rather than machine learning.
We test enacting cuts based on simple data-driven photometric variability parameters to distinguish non-AGN extragalactic transients from standard AGN variability on both Zwicky Transient Facility photometry and simulated LSST photometry from the MALLORN data set. We also investigate the impact of light curve history availability, redshift range and filter selection on selection efficiency. We find that a two-dimensional cut incorporating the ratio of detection flux and pre-detection standard deviation and the ratio of detection flux to pre-detection mean flux is the most effective cut. This approach is easily scalable as these values are included in the LSST alert packets.
We provide estimates of the completeness and purity of the sample produced by enacting this cut, and gauge the AGN contamination avoided. The parameters utilised in this approach could also be implemented as features for identifying AGN in a photometric classifier.

\end{abstract}




\section{Introduction}

The Vera C. Rubin Observatory's 10-Year Legacy Survey of Space and Time \citep[LSST;][]{LSST} is expected to produce a deluge of new transient detections. Thanks to its large collecting area (8.4\,m primary mirror), it can observe to greater depths ($r_{lim}$ = 24.5) than previous wide-field survey telescopes. LSST is expected to produce 10 million nightly transient alerts, two orders of magnitude greater than the number currently produced by the Zwicky Transient Facility \citep[ZTF;][]{ZTF}. Through these alerts, LSST is predicted to discover \textasciitilde10 million supernovae (SNe) during the survey, corresponding to approximately a million SNe discovered per year.

Dedicated spectroscopic follow-up on this wealth of transient detections will be provided by several facilities, including the European Southern Observatory (ESO) 4-m Multi-Object Spectroscopic Telescope (4MOST), conducting the Time Domain Extragalactic Survey \citep[TiDES;][]{TiDES}, the Son of X-shooter \citep[SOXS;][]{SOXS} instrument on the New Technology Telescope (NTT), the Rubin-ESO Investigative Spectroscopy Of Target Transient Objects (RISOTTO) survey on the ESO Very Large Telescope (VLT), and US NOIRLab telescopes such as Gemini \citep[][]{Gemini} and SOAR \citep[][]{SOAR}. Despite the efforts of this dedicated follow-up, we will only be able to get spectra for a few percent of the objects that LSST will discover. Therefore, it is important that we are able to allocate our limited resources effectively to maximise scientific return.

The data from LSST is useful for several transient science goals. LSST is expected to discover 400,000 photometrically classified SNe Ia with multi-band light curves useful for cosmological distance measurements \citep[][]{LSST}, resolving current tensions in the nature of dark energy \citep[][]{AbdulKarim2025,Popovic2026}. A subset of these will receive spectroscopic follow-up of the SN and/or the host galaxy, further enabling statistical studies of rates and progenitor scenarios \citep[e.g.,][]{Wiseman2021,Srivastav2026} and unaccounted for astrophysical cosmological systematics \citep[e.g.,][]{Toy2025,Ramaiya2025,Burgaz2026}. This includes the potential for discovery of higher redshift SNe Ia, which enable improved constraints on the properties of dark energy as a function of redshift \citep[][]{Garnavich2004}. The depth and consistent cadence of LSST is predicted to discover progenitors for core-collapse supernovae (CCSNe), thereby improving our understanding of progenitor properties, potentially answering the `Red Supergiant Problem' \citep[][]{RSGProblemSmartt} and improving our progenitor modelling capabilities. LSST has the potential to serendipitously observe fast explosions with short durations, such as luminous fast blue optical transients \citep[LFBOTs;][]{Drout2014, Prentice2018, Pursiainen2018, Perley2019, Ho2020, Chrimes2024, Gutierrez2024}, eclipses in ultracompact double-degenerate binary systems \citep[][]{Anderson2005} and luminous fast coolers \citep[LFCs;][]{Nicholl2023}, as well as explosions with unusually long durations such as pair instability supernovae \citep[PISNe;][]{Barkat1967,Rakavy1967,Bond1984,Heger2002,Moriya2010,Kasen2011}. Extreme transients from supermassive black holes (SMBH) will also be investigated by LSST, including tidal disruption events \citep[TDEs;][]{vanVelzen2011,Gezari2012,Gezari2021} and ambiguous nuclear transients \citep[ANTs;][]{Kankare2017, Wiseman2025, Hinkle2025}, greatly enhancing the sample size for these transients and enabling many new avenues of research \citep[][]{vanVelzen2011,Bricman2020,BucarBricman2023}. The Rubin Observatory also hopes to identify 500-1000 multiply imaged strongly lensed SNe Ia \citep[][]{Goldstein2017} and several hundred strongly lensed CCSNe \citep[][]{Oguri2010,Coulter2026}. The time delays between images allow for extensive observation of the shock breakout phase of the light curve, probing the early emission from CCSNe \citep[][]{Suwa2018}. LSST is anticipated to produce a large sample of superluminous supernovae (SLSNe) with the potential for cosmological applications \citep[][]{Scovacricchi2016,Inserra2021}. The Rubin Observatory also aims to identify optical counterparts to multi-messenger sources such as kilonovae \citep[KNe;][]{Abbott2017,Valenti2017,Smartt2017,Pian2017,Scolnic2018}, gamma ray bursts \citep[GRBs;][]{Sahu1997,Zhang2004,Zhang2006,Kann2010} and neutrino events detected by \texttt{ICECUBE}\footnote{\url{https://icecube.wisc.edu/}} \citep[][]{Stein2021,vanVelzen2024,Toshikage2025}. However, these transients of interest to the LSST science goals are not the only things that vary in brightness, particularly in galaxy centres.

Of key concern is the potential for previously uncatalogued active galactic nuclei (AGN) to produce an avalanche of alerts. LSST will likely detect many new AGN not present in any existing catalogues, such as \texttt{Milliquas} \citep[][]{milliquas}. If the luminosity of an AGN is near the LSST detection limits, the stochastic variation in the AGN light curve may cause it to peak above and dip below the detection thresholds, each time producing an alert which, if not accounted for, could contaminate the target selection for LSST follow-up surveys -- spending valuable observing time on unintended targets. Whilst AGN are a priority science target for LSST - with the sample of variability selected AGN expected to increase by a factor of 4, providing important constraints on black hole demographics \citep[][]{Kaviraj2026} - their persistent nature, as opposed to the explosive, brief nature of other transients, means they do not require the rapid classification and follow-up spectra necessary for transient science. This is of particular concern for surveys such as TiDES with automatic target selection and triggering.


As matter accretes onto a black hole, the outward transfer of angular momentum results in the formation of a luminous accretion disk visible to an observer as an AGN \citep[][]{Prendergast1960, LyndenBell1969, Shakura1972, PringleRees1972}. We do not have a complete physical theory of AGN variability, but there are several processes in the disk that could potentially contribute. For example, variability can be produced by autoregulation of the accretion disk, due to coupling between the rate of accretion, the resulting accretion luminosity, and the radiation pressure in the disk. Chaotic small-scale magnetic reconnections caused by differential rotation in the accretion disk and turbulence effects can also influence the luminosity variability \citep[][]{ShakuraSunyaev1973, Rees1984, Ulrich1997, Kelly2009, Kelly2014}. Aspects of this variability are defined as either stochastic \citep[][]{Shvartsman1971} or quasi-periodic \citep[][]{Sunyaev1972}, and can be modelled as a damped random walk \citep[][]{MacLeod2010}.

This variability inherent to AGN has been a point of interest for researchers since their discovery in the 1960s \citep[][]{Greenstein1963, Hazard1963, Matthews1963, Oke1963, Schmidt1963}. Observations of this variability can be used to infer the structure and kinematics of the broad-line regions of AGN via reverberation mapping, allowing for an estimate of the central black hole mass \citep[][]{Peterson2004, ReverbMapping}. Large numbers of AGN have been identified via wide-field surveys, allowing for research into this phenomenon with data from the Sloan Digital Sky Survey \citep[SDSS;][]{VandenBerk2004, Wilhite2005, Branimir2007}, Zwicky Transient Facility \citep[ZTF;][]{Jha2022} and the Asteroid Terrestrial Last Alert System \citep[ATLAS;][]{Tan2026}. However, for surveys focused on explosive transient discovery (i.e. that of SNe, TDEs, etc.), these objects are often regarded as unwanted contaminants in the search for other transients \citep[][]{ATLAS_VRA}.

To separate AGN from non-AGN, a mid-infrared colour criterion of \textit{Wide-field Infrared Survey Explorer} \citep[\textit{WISE};][]{WISE} bands $W1$ ($3.4\mu m$) - $W2$ ($4.6\mu m$) $>$ 0.8 can be used. Using this approach identifies 62 AGN per deg$^2$ to a depth of $W2$ = 18.3 (AB mag). This method is capable of identifying both Type 1 (unobscured) and Type 2 (obscured) AGN with a reliability of 95\% \citep[][]{WISEAGN}. This works by distinguishing the power-law AGN spectrum from the luminosity-weighted stellar spectra that galaxies are comprised of. The effectiveness of this approach was predicted in \citet{Ashby2009}, \citet{Assef2010} and \citet{Eckart2010}, and subsequently demonstrated in \citet{WISEAGN}. Transient surveys often use \textit{WISE} colour cuts to reject likely AGN. Whilst this approach can successfully identify some high-redshift sources, there is a large mismatch between the sensitivity of LSST (reaching $\sim24.5$\,AB mag in the optical) and \textit{WISE} (AB limiting magnitudes $W1$ = 19.2, $W2$ = 18.9), thus a large quantity of LSST AGN will not be present in the \textit{WISE} catalog (we show this quantitatively in Appendix \ref{APP_WISE}). Consequently, it is necessary for one to consider alternative methods for separating AGN detections from those of other transients.

There are many machine learning codes which aim to determine the likely physical type of a transient based on its available photometry and host information. These can be dedicated to the identification of a particular transient; for example, codes such as NEEDLE \citep[][]{Sheng2024}, TDEscore \citep[][]{Stein2024} and FLEET \citep[][]{Gomez2023} are designed to identify TDEs and/or SLSNe. Other classifiers are more general purpose, aiming for photometric classification of many transient types; such codes include the ALeRCE Light Curve Classifier \citep[][]{SanchezSaez2021}, SuperRAENN \citep[][]{Villar2019,Villar2020}, and SuperNNova \citep[][]{Moller2016,Moller2020,Moller2025}. To distinguish AGN from other transients, these classifiers often make use of existing AGN catalogues, photometric variability parameters calculated from the light curve, and host properties including the aforementioned WISE colours.

However, the use of machine learning can make it non-trivial to characterise the selection function, as these algorithms are not always fully interpretable. This can introduce complex biases in calculations of demographics and rates if the classifier isn't perfectly understood. For surveys such as TiDES, which aim to produce a well-understood statistical sample, a simpler selection strategy that can be easily modelled and reproduced is necessary to estimate the completeness and purity of the sample. The accuracy of these values is vital for the scientific utility of such data sets, particularly for cosmological applications and transient rate studies.

In this paper, we design a set of possible selection strategies based only on data-driven cuts on photometric variability using no prior information from catalogues or host galaxies. We evaluate the effectiveness of implementing these strategies to separate AGN from other transients. We assess a variety of different parameters and combinations of parameters on two data sets - a Zwicky Transient Facility \citep[ZTF;][]{ZTF} nuclear photometric data set and the MALLORN \citep[][]{MALLORN} simulated LSST photometric data set. In Section \ref{Section2}, we outline the data sets used, explain each parameter cut and evaluate their effectiveness at distinguishing non-AGN transients from AGN in ZTF photometry. In Section \ref{Section3}, we create a multivariable cut which is optimised for distinguishing non-AGN from AGN in LSST and evaluate the effectiveness of this cut on ZTF photometry. We also investigate the impact of light curve history availability and redshift. We perform a test of this approach on an unlabelled ZTF sample to validate its performance on new data. In Section \ref{Section4}, we repeat the above mentioned tests on MALLORN simulated LSST data to estimate the performance of the same transient selection strategies on LSST-like light curves. In Section \ref{Section5}, we reiterate our overall findings and provide an estimate of the fibre-hours saved by TiDES by implementing our suggested multivariable approach. The code relevant to this paper can be accessed via GitHub\footnote{\url{https://github.com/dkjmagill/AGN_Phot_Var_Screening}}.

\section{Photometric Variability Parameters}
\label{Section2}

\subsection{Data sets used}
To identify strategies to differentiate explosive transients from AGN, we use two data sets to test cuts on various photometric parameters.

The ZTF data consists of all of the ZTF forced photometry of nuclear transients from 2017-2023. A `Nuclear Filter' originally written and used by \citet{VanVelzen2019,VanVelzen2021,Hammerstein2021} to select nuclear transients in real-time from the ZTF alert stream was applied to the \texttt{AMPEL} archive of ZTF alerts \citep{AMPEL}. To pass the `Nuclear Filter' an object requires a high likelihood of a galactic host, at least one detection brighter than 20 mag and an angular distance to the core of less than 0.5 arcseconds. For more details on this filter, please see \citep[][]{Reusch2024}. The sample data set produced by this filter consisted of 11,190 objects. There are no hostless transients present in the sample.

The ZTF forced photometry pipeline was run for all transients in our sample, returning the time, count rate and zeropoint for each observation in the ZTF $g$ and $r$ bands. The zeropoint is used to convert each observation to a flux density measurement in microjanksys ($\mu$Jy). 
In this section we show only the results calculated on the ZTF $g$ band data. For the same tests on the ZTF $r$ band data, which yield equivalent results, see Appendix \ref{APP_ZTF_r}.

Via a crossmatch with the \texttt{Transient Name Server}\footnote{\hyperlink{https://www.wis-tns.org/}{\texttt{Transient Name Server}}}, the spectroscopic type and redshift were determined for 791 objects, with a further 38 found via manual crossmatching. Further crossmatching with version 8 of \texttt{Milliquas} identified 1407 additional AGN present within the ZTF data set, bringing the total number of labelled objects to 2236. Objects misclassified as Type II SLSNe on \texttt{TNS} were corrected to AGN based on the contaminants identified in \citet{Pessi2025}. Transient candidate selection processes are likely biased against transients, most notably NIR TDEs, in very luminous and dusty environments \citep[][]{Masterson2024}. More work is required for a thorough understanding of these objects, and consequently it is possible that there are some of these objects included in the sample which are unaccounted for.

Since many objects still lack spectroscopic classifications, we further split the data into a labelled set, in which the type and redshift of the objects were known, and a blind data set for which there is no known type or redshift. The labelled data set allowed for the design and evaluation of different transient selection strategies, and the calculation of completeness, purity and F1 score values. We used the blind sample to test our final chosen approach on a real sample of transients of unclear type, allowing for the effectiveness of the cuts on new data to be assessed.

The Many Artificial LSST Light curves based on Observations of Real Nuclear transients (MALLORN) data set was used to evaluate the utility of the same transient selection strategies on LSST-like light curves. MALLORN consists of 10,178 simulated LSST light curves of various types of transients (SNe, TDEs and AGN). The data set was constructed from the ZTF nuclear sample and consequently the MALLORN light curves are firmly grounded in real observational data. For more information on the creation and contents of the MALLORN data set, please see \citet{MALLORN}.

\subsection{Parameters and statistics used}

Photometric variability is a simple but promising means to separate AGN from explosive transients since we expect AGN to vary stochastically over the duration of a survey, whereas we expect explosive transients to appear suddenly, usually with no prior flux variation at the transient position. However, in some cases, this divide is not necessarily so clear-cut. Type II SNe can exhibit pre-explosion variability produced by progenitor luminosity variability and mass loss \citep[][]{Schmidt1992, Pastorello2007, Smartt2009, Goldberg2026, Chen2026}; however, with respect to the detection flux, this variability is fainter by orders of magnitude, and consequently, these transients are still very distinct from standard AGN activity. TDEs and SNe can also occur in galaxies that already harbour AGN, and therefore appear to show prior variability even if this is not related to the transient itself. In such cases, large increases in flux caused by the transient are still of interest to transient astronomers. Furthermore, AGN occasionally show large, coherent and sustained increases in brightness \citep[e.g.][]{Lawrence2012,Graham2017,Kankare2017,Trakhtenbrot2019,Petrushevska2023,Wiseman2025,Graham2026} that are not well modelled as stochastic variability, and follow-up observations to understand the physics of such unusual modes of AGN variability is desirable. For these reasons, transient selection is not as straightforward as rejecting any event that shows prior variability.

The following light curve statistics were evaluated for their utility in being used to separate AGN from explosive transients: root mean square (RMS) flux, standard deviation, mean flux (all calculated on the long-term difference light curve at the transient position, prior to transient discovery) and flux contrast of the detected transient in difference imaging compared to the host nucleus in archival imaging.
We define transient discovery as the first 5$\sigma$ detection in the selected wavelength band. 

To evaluate the effectiveness of selecting transients using each parameter described in the following sub-sections, we assess the resultant completeness, purity and F1 scores. Completeness, defined as the number of true positives (correct positive predictions) relative to the total expected number of positives, provides the fraction of the desired transients which pass the cut. Purity, defined by the number of true positives over the sum of true positives and false positives (incorrect positive predictions), illustrates how many undesired objects are passing the cut (a higher purity means fewer contaminants). Ideally, we want to maximise both of these values to produce a clean and full sample; however, there is often a trade-off: enacting a more strict cut will enhance the purity of the sample, but may diminish its completeness.

The F1 score provides a means to optimise the middle-ground. It is the harmonic mean of the completeness and purity, with an F1 score of 1 corresponding to the best possible performance and a score of 0 to the worst. The completeness and purity values have equal weighting in the calculation of the F1 score, which is calculated using
\begin{equation}
    \mathrm{F} 1=\frac{2 \times \mathrm{TP}}{2 \times \mathrm{TP}+\mathrm{FP}+\mathrm{FN}},
\end{equation}
where TP is the number of true positives , FN is the number of false negatives (incorrect negative predictions) and FP is the number of false positives .

\subsection{Historical Standard Deviation}
\label{Section2.3}
Standard deviation ($\sigma$) is a statistical measure of the extent to which a data set varies from its average value. A low standard deviation indicates that all values are quite close to the mean value, whereas a high standard deviation value demonstrates that values in the data set are frequently significantly different to the mean value and hence suggests that the data is more variable.

\begin{figure}
	\includegraphics[width=\columnwidth]{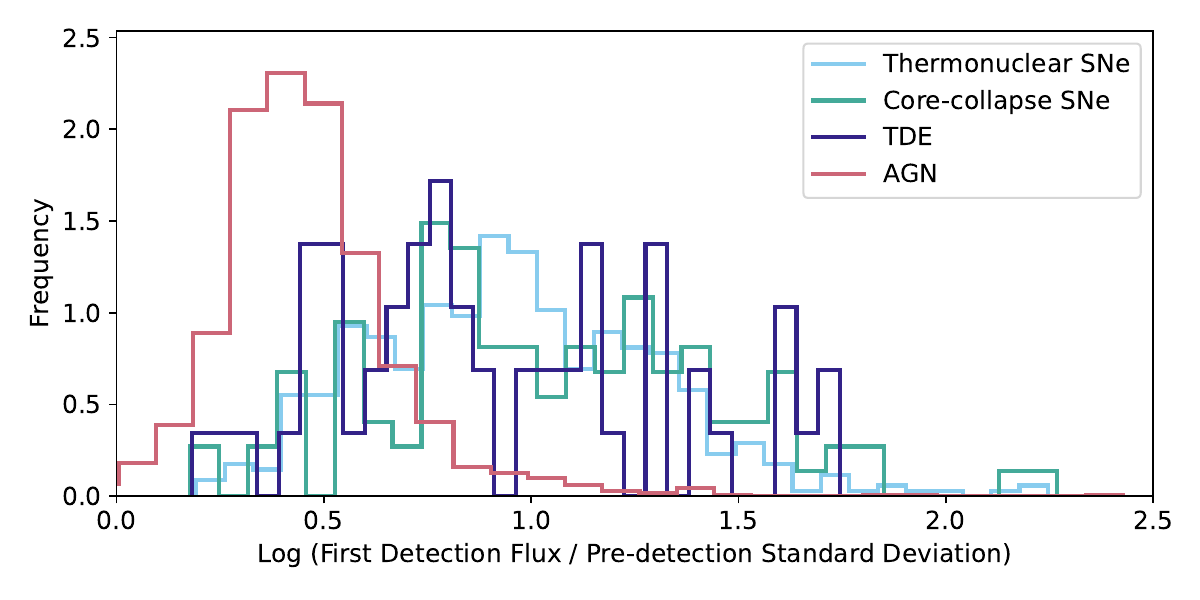}
    \caption{Histogram showing the distribution of the ratio of the $g$-band first detection flux to pre-detection standard deviation for thermonuclear supernovae, core-collapse supernovae, tidal disruption events and active galactic nuclei. There is a clear separation between the peaks of the AGN and transient distributions.}
    \label{fig:std_hist}
\end{figure}

We would therefore expect the pre-transient flux for a quiescent transient host galaxy to have a low standard deviation, while an AGN will have more variability, producing a larger standard deviation. To avoid rejecting transients that are genuine explosions but are coincident with some previous variability (e.g. from the progenitor or more likely due to a spatially coincident AGN), we calculate the ratio of first detection flux to the pre-detection standard deviation. The validity of this approach is demonstrated in Figure \ref{fig:std_hist}. There is a clear separation between the peak of the AGN distribution (\textasciitilde 0.4) and that of the other transient types (\textasciitilde 0.9).

This matches our intuition: transients in quiescent galaxies will show a large contrast, whereas standard AGN stochastic variability will produce a low contrast given their greater standard deviation. For genuine transients of interest that happen to occur within an AGN, we can find the value of this ratio that maximises the F1 score.

\begin{figure}
	\includegraphics[width=\columnwidth]{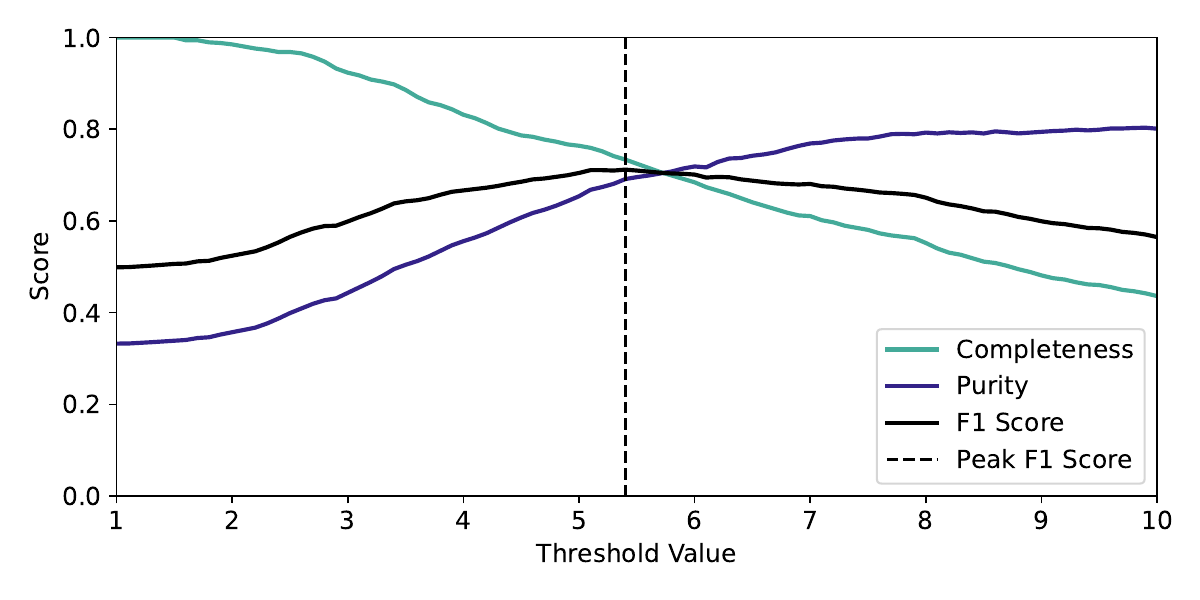}
    \caption{Plot of completeness, purity and F1 score against parameter threshold value for the $g$ band ratio of detection flux to pre-detection flux standard deviation. The F1 score peaks at a value of 0.71 at $F_{\rm det} / \sigma_F > 5.4$ (indicated by the dashed line), with corresponding completeness and purity values of 0.73 and 0.69, respectively.}
    
    \label{fig:std_cpf1}
\end{figure}

We vary the threshold ratio of detection flux to pre-detection standard deviation, $F_{\rm det}/\sigma_F$, where $F_{\rm det}$ is the detection flux, between 1 and 10 in steps of 0.1. At each step, we determine which of the classified ZTF transient and AGN light curves pass the cut and which fail. Using this we can calculate the corresponding completeness, purity, and F1 scores for each cut, as shown in Figure \ref{fig:std_cpf1}. 

For this parameter, a peak F1 score of 0.71 is attained at $F_{\rm det} / \sigma_F > 5.4$. The corresponding completeness and purity values produced are 0.73 and 0.69, respectively. 

\subsection{Historical Mean Flux}
\label{Section2.4}
\begin{figure}
	\includegraphics[width=\columnwidth]{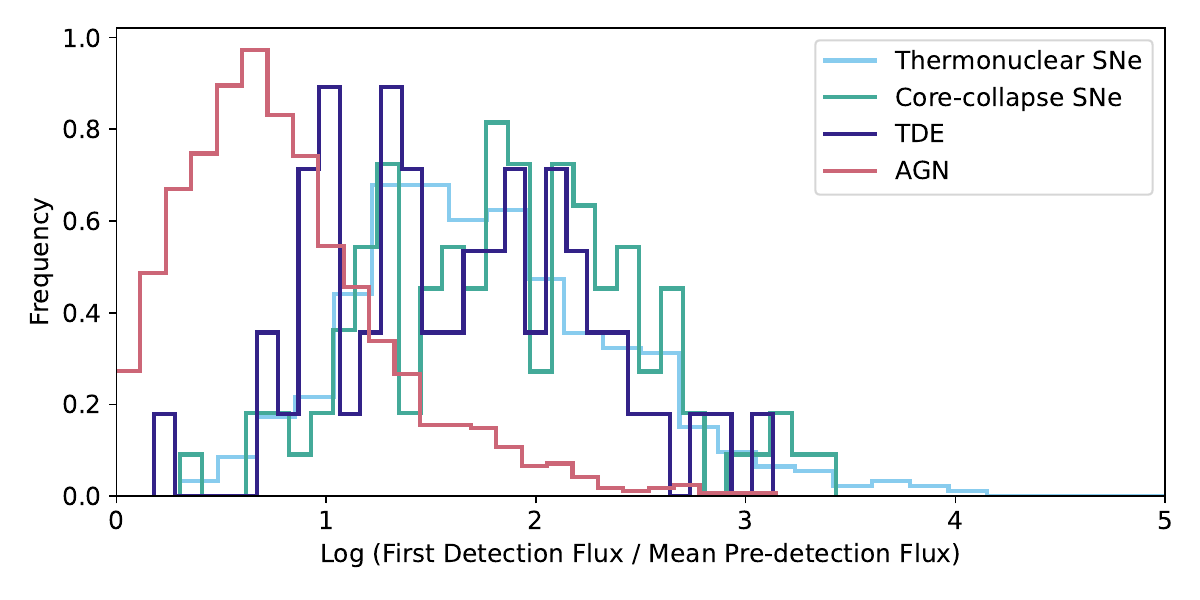}
    \caption{Histogram showing the $g$ band distribution of the log ratio of first detection flux to pre-detection mean flux for thermonuclear supernovae, core-collapse supernovae, tidal disruption events and active galactic nuclei. There is a clear separation between the peaks of the AGN and transient distributions.}
    \label{fig:mf_hist}
\end{figure}

The ratio of the detection flux to the mean of the pre-detection flux, $\langle F \rangle$, can also serve as a method of distinguishing transients of interest from AGN variability. An explosive transient with no prior variability will have $\langle F \rangle$ close to zero before detection, whereas any previous epochs of AGN flaring will inflate the mean of the historic light curve. The validity of this approach is demonstrated by Figure \ref{fig:mf_hist}. There is a clear separation between the peak of the AGN distribution (\textasciitilde 0.7) and that of the other transient types (\textasciitilde 1.9), indicating that this parameter has the potential to be used to distinguish AGN from other transients.

\begin{figure}
	\includegraphics[width=\columnwidth]{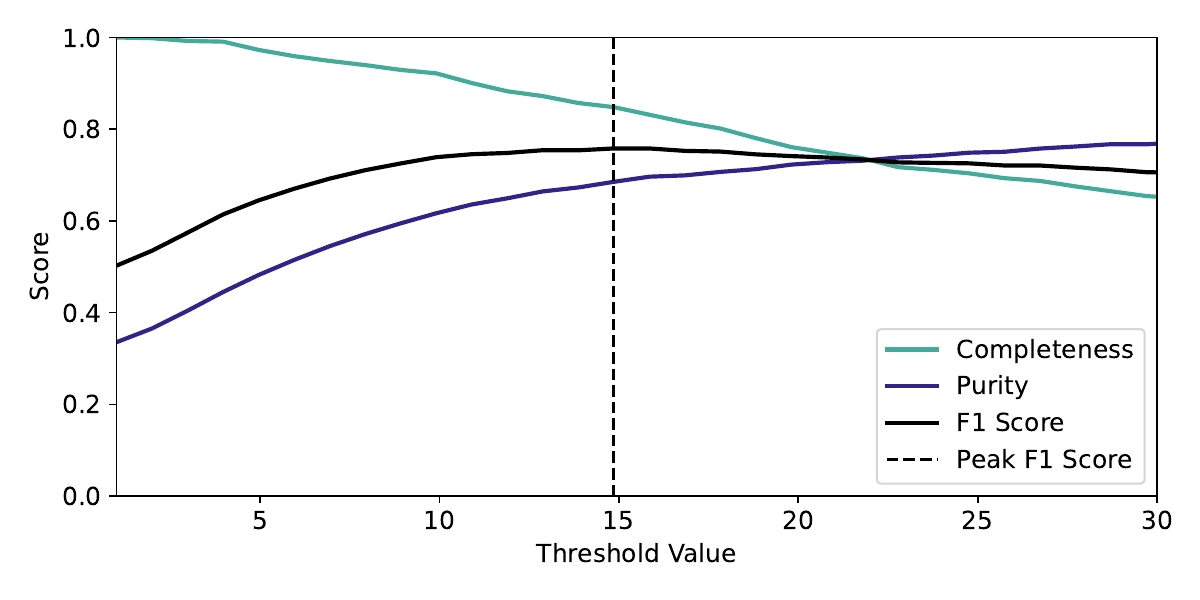}
    \caption{Plot of completeness, purity and F1 score against parameter threshold value for the $g$ band ratio of detection flux to pre-detection flux mean. The F1 score peaks at a value of 0.76 at $F_{\rm det} / \langle F \rangle > 14.9$ (indicated by the dashed line), with corresponding completeness and purity values of 0.85 and 0.69, respectively.}
    \label{fig:mf_cpf1}
\end{figure}

We vary the threshold ratio of detection flux to pre-detection mean flux, $F_{\rm det}/\langle F \rangle$, between 1 and 30 in steps of 0.1. At each step, we determine which of the classified ZTF transient and AGN light curves pass the cut and which fail. Using this we can calculate the corresponding completeness, purity, and F1 scores for each cut, as shown in Figure \ref{fig:mf_cpf1}.

A peak F1 score of 0.76 is attained at $F_{\rm det} / \langle F \rangle > 14.9$. The corresponding completeness and purity values produced are 0.85 and 0.69, respectively. $F_{\rm det} / \langle F \rangle$ achieves a higher peak F1 score than $F_{\rm det} / \sigma_F$. This is due to it producing significantly higher completeness while the purity remains the same at the peak F1 score.

\subsection{Root Mean Square Variation}
\label{Section2.5}
\begin{figure}
	\includegraphics[width=\columnwidth]{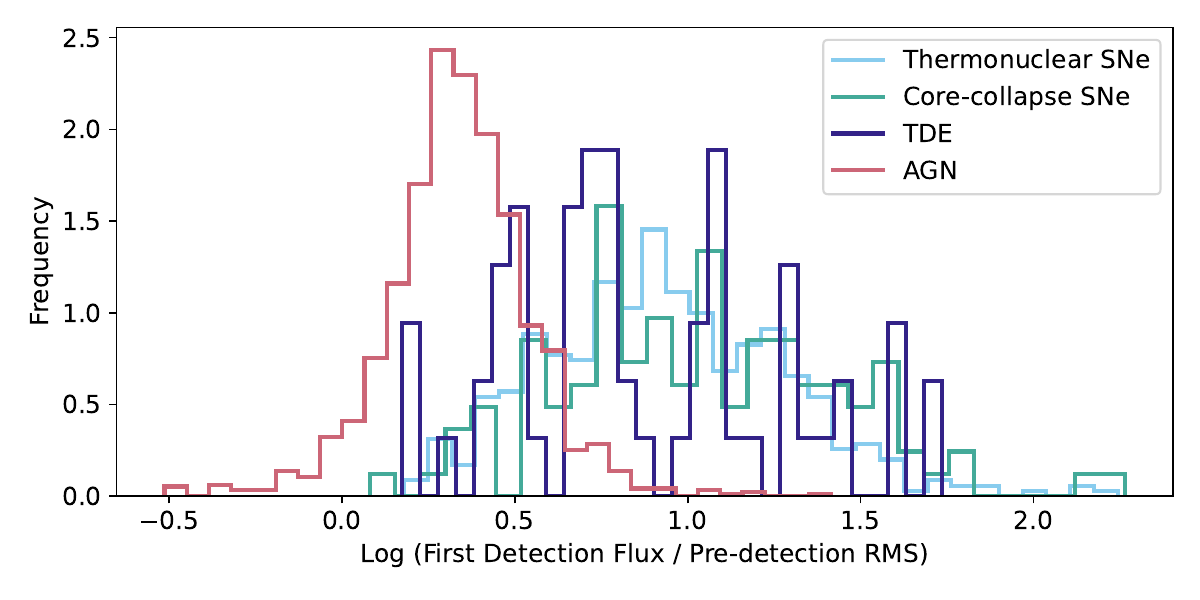}
    \caption{Histogram showing the $g$ band distribution of the log ratio of first detection flux to pre-detection root mean square for thermonuclear supernovae, core-collapse supernovae, tidal disruption events and active galactic nuclei. There is a clear separation between the peaks of the AGN and transient distributions.}
    \label{fig:rms_hist}
\end{figure}

The root mean square (RMS) is an alternative way to parametrise the variability of a data set. It is calculated by the following formula:
\begin{equation}
    \mathrm{RMS_F} = \sqrt{\frac{\sum_{i=1}^{N} (F_i)}{N}},
    \label{eq:rms_equation}
\end{equation}
where $F_i$ refers to a given data point and $N$ refers to the number of data points. We expect standard stochastic AGN variability to produce a larger RMS value, whilst a quiescent host will have little variability and hence a smaller RMS value. Consequently, and in a similar manner to the mean and standard deviation, by using the ratio between detection flux and pre-detection RMS, we can impose a cut to distinguish AGN variability from other transients. This is demonstrated in Figure \ref{fig:rms_hist}. There is a clear separation between the peak of the AGN distribution (\textasciitilde 0.25) and that of the other transient types (\textasciitilde 0.9), supporting the use of this parameter as a metric to distinguish AGN from other transients.

\begin{figure}
	\includegraphics[width=\columnwidth]{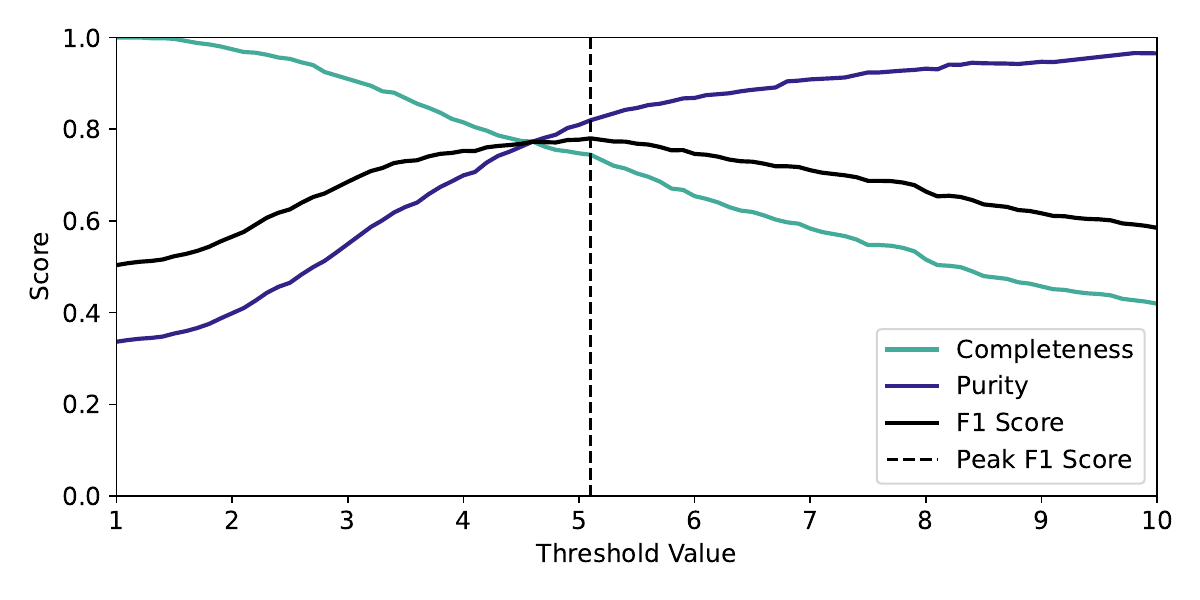}
    \caption{Plot of completeness, purity and F1 score against parameter threshold value for the $g$ band ratio of detection flux to pre-detection flux RMS. The F1 score peaks at a value of 0.78 at $F_{\rm det} / RMS_F > 5.1$ (indicated by the dashed line), with corresponding completeness and purity values of 0.74 and 0.82, respectively.}
    \label{fig:rms_cpf1}
\end{figure}

We vary the threshold ratio of detection flux to pre-detection RMS, $F_{\rm det} / RMS_F$, between 1 and 10 in steps of 0.1. At each step, we determine which of the classified ZTF transient and AGN light curves pass the cut and which fail. Using this we can calculate the corresponding completeness, purity, and F1 scores for each cut, as shown in Figure \ref{fig:rms_cpf1}. 

A peak F1 score of 0.78 is attained at $F_{\rm det} / RMS_F > 5.1$. The corresponding completeness and purity values produced are 0.74 and 0.82, respectively. The peak F1 score produced by this approach is greater than the two previous methods, as it achieves a greater completeness and a significantly greater purity at peak F1 score.

A similar approach was employed in \citet{vanVelzen2024}, in which $\Delta F_{\rm IR} / RMS_F$ was applied to infrared observations to determine dust echo strength from nuclear transients. Selected candidate dust echoes required echo strength larger than the significance of the baseline RMS. Though applied at different wavelengths, this approach is similar in principle and supports the use of these metrics to distinguish transients of significant strength from standard variability.

\subsection{Host Contrast}
Another potential method of distinguishing AGN from other transients is to compare the detection flux to prior measurements of the host nucleus.  Whereas our other methods are technically agnostic to the transient location (though likely only applicable to nuclear transients, since off-nuclear transients are overwhelmingly supernovae - although see \citet{Ward2021}, \citet{Yao2025} \& \citet{Stein2026}), this approach is specific to nuclear transients coincident with a catalogued galaxy. The method we employed is based on that described in Section 3.5 of \citet{Hung2018}. Specifically, we compute:

\begin{equation}
\begin{aligned}
    \Delta m_{\mathrm{var}} &= m_{\rm PSF,sci} - m_{\mathrm{PSF,host}}\\
            &= -2.5 \log_{10}\left( F_{\rm PSF,host} + F \right) - m_{\mathrm{PSF,host}},
\end{aligned}
\end{equation}
where $\Delta m_{\mathrm{var}}$ refers to the amplitude of the variability and $m_{\mathrm{PSF,host}}$ refers to the catalogued point spread function (PSF) magnitude of the host. If flux measurements are available on the un-differenced transient detection image (i.e. the science image with no reference subtracted) then the science magnitude $m_{\rm PSF,sci}$ can be used directly. Otherwise, this can be computed from the transient flux $F$ in the difference image and the host flux $F_{\mathrm{PSF,host}}$ (which is related to $m_{\mathrm{PSF,host}}$ in the usual way). The PSF magnitude is used even though galaxies are not typically well described as point sources, since the PSF fit primarily captures light only from the nucleus.

\begin{figure}
	\includegraphics[width=\columnwidth]{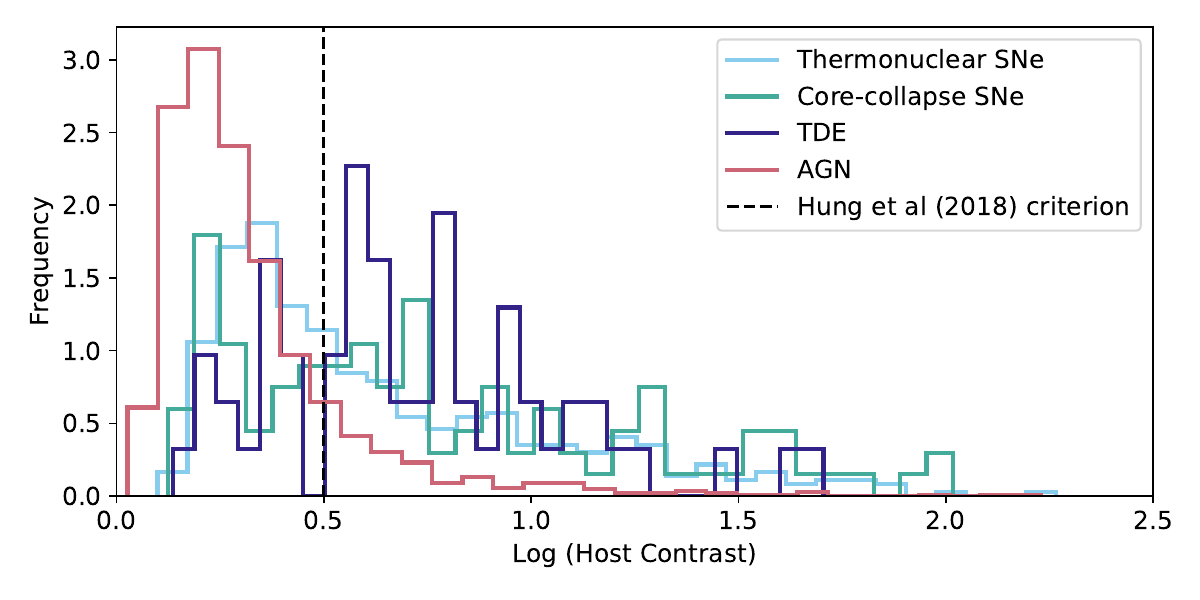}
    \caption{Histogram showing the $g$ band distribution of the log of the host contrast parameter defined in \citet{Hung2018} for thermonuclear supernovae, core-collapse supernovae, tidal disruption events and active galactic nuclei. There is a clear separation between the peaks of the AGN and transient distributions. The threshold value of 0.5 implemented in \citet{Hung2018} is indicated with a dashed black line.}
    \label{fig:host_contrast_hist}
\end{figure}

We source the host magnitudes for each transient in the labelled ZTF data set from the 3pi Pan-STARRS catalogue \citep[][]{PS1Cat}. Normal stochastic AGN variability will typically have $\Delta m_{\mathrm{var}} < 0.5$, whereas explosive transients and high-amplitude flares of interest should have larger values. The validity of using this approach is demonstrated in Figure \ref{fig:host_contrast_hist}. There is a clear separation between the peak of the AGN distribution (\textasciitilde 0.2) and that of the other transient types (\textasciitilde 0.6), thereby indicating that this parameter can be used to distinguish AGN variability from other transients.

\begin{figure}
	\includegraphics[width=\columnwidth]{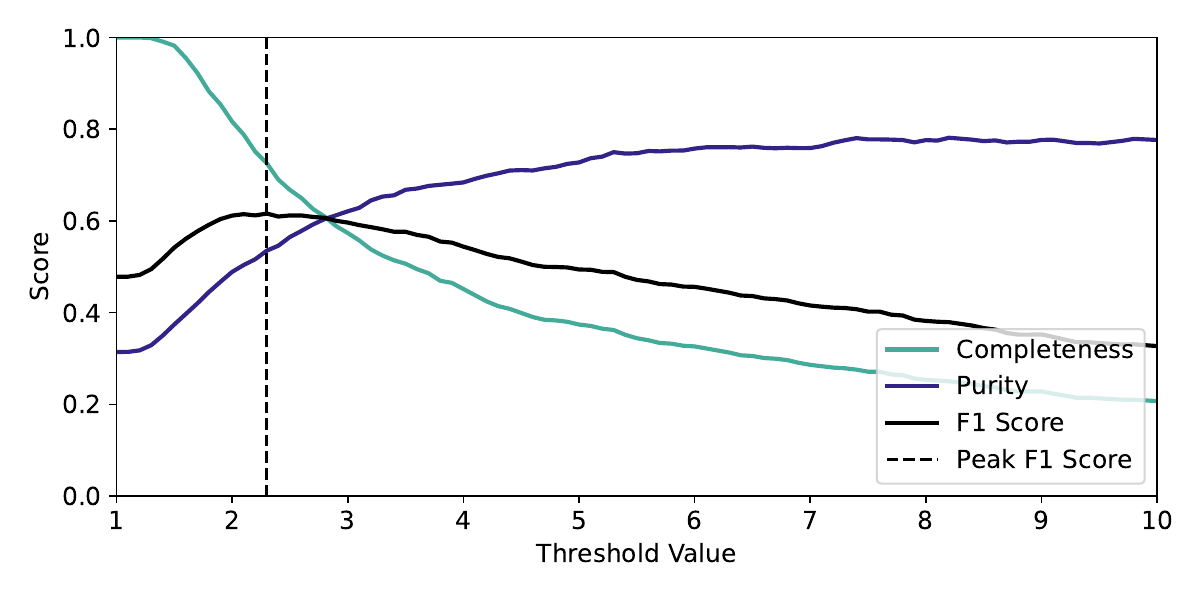}
    \caption{Plot of completeness, purity and F1 score against host contrast parameter threshold value. The F1 score peaks at a value of 0.62 at $\Delta m_{\mathrm{var}} > 2.3$. (indicated by the dashed line), with corresponding completeness and purity values of 0.73 and 0.54, respectively.}
    \label{fig:hc_cpf1}
\end{figure}

We vary the threshold ratio of host contrast, $\Delta m_{\mathrm{var}}$, between 1 and 10 in steps of 0.1. At each step, we determine which of the classified ZTF transient and AGN light curves pass the cut and which fail. Using this we can calculate the corresponding completeness, purity and F1 scores for each cut, as shown in Figure \ref{fig:hc_cpf1}. A peak F1 score of 0.62 is attained at $\Delta m_{\mathrm{var}} > 2.3$. The corresponding completeness and purity values at this threshold value are 0.73 and 0.54, respectively. This F1 score is lower than those produced by the previous methods, largely due to a much lower purity value, while the completeness remains comparable. 

The results of these tests are shown in Table \ref{tab:ztf_results_table}. $F_{\rm det} / RMS_F > 5.1$ produces the best F1 score, with high completeness and very high purity. $F_{\rm det} / \langle F \rangle > 14.9$ produces the next best F1 score, with greater completeness than $F_{\rm det} / RMS_F$ but significantly lower purity. $F_{\rm det} / \sigma_F > 5.4$ produces the third best F1 score of the parameters tested, with purity comparable to $F_{\rm det} / \langle F \rangle$ but lower completeness comparable to that of $F_{\rm det} / RMS_F$. $\Delta m_{\mathrm{var}} > 2.3$ produces the lowest F1 score, despite completeness comparable to $F_{\rm det} / \sigma_F$ and $F_{\rm det} / RMS_F$, it has a much lower purity score which significantly reduces its F1 score.

\begin{table}
    \centering
    \caption{Results table showing effectiveness of each parameter applied to the ZTF nuclear sample.}
    \label{tab:ztf_results_table}
    \begin{tabular}{lllll}
        \hline
        Parameter & Threshold & F1 & Completeness & Purity \\
        \hline
        $F_{\rm det} / \sigma_F$ & 5.4 & 0.71 & 0.73 & 0.69 \\
        $F_{\rm det} / \langle F \rangle$ & 14.9 & 0.76 & 0.85 & 0.69 \\
        $F_{\rm det} / RMS_F$ & 5.1 & 0.78 & 0.74 & 0.82 \\
        $\Delta m_{\mathrm{var}}$ & 2.3 & 0.62 & 0.73 & 0.54 \\
        \hline
    \end{tabular}
\end{table}

\section{Designing the Optimal Cut for LSST}
\label{Section3}

\subsection{Implementing a Multivariable Cut}
\label{Section3.1}

\begin{figure*}
    \centering
    \includegraphics[width=\textwidth]{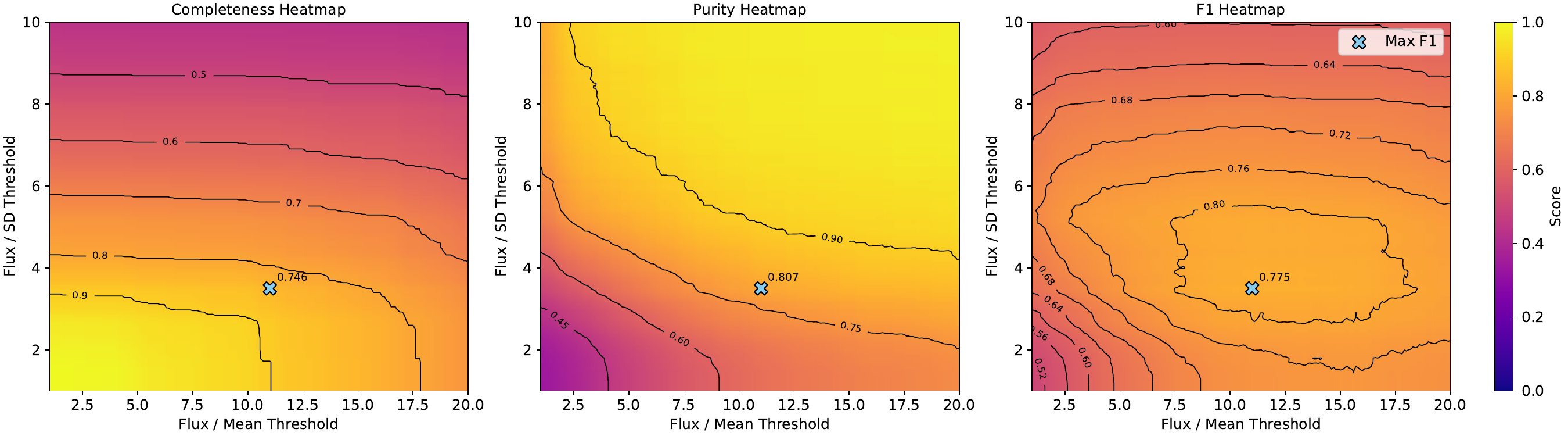}
    \caption{Heatmaps indicating the completeness (left), purity (centre) and F1 score produced by a range of cuts on the ZTF nuclear data set for $F_{\rm det}/\sigma_F$ (y-axis) and $F_{\rm det}/\langle F \rangle$ (x-axis). The peak F1 score value (indicated by the marker) occurs at a value of 0.78, with corresponding completeness and purity values of 0.75 and 0.81, respectively.}
    \label{fig:ztf_2d}
\end{figure*}

\begin{figure}
	\includegraphics[width=\columnwidth]{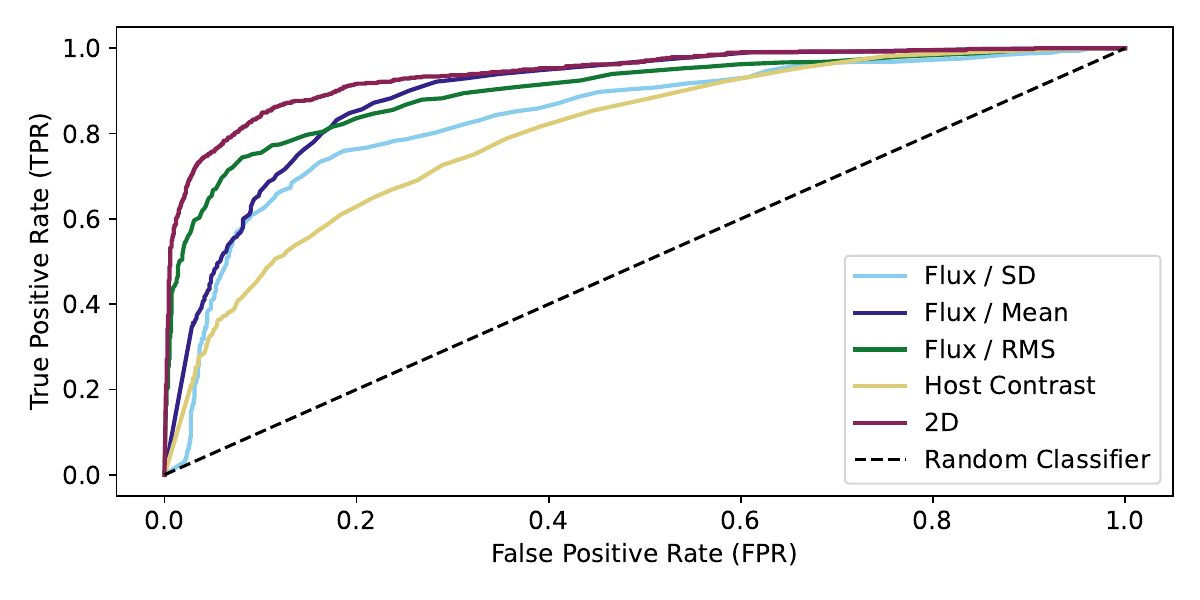}
    \caption{$g$ band ROC curves plotting the true positive rate against the false positive rate for the following parameter cuts applied to the labelled ZTF nuclear data set: the ratio of detection flux to pre-detection standard deviation, the ratio of detection flux to pre-detection mean flux, the ratio of detection flux to pre-detection root mean square, the host contrast parameter from \citet{Hung2018} and the two-dimensional cut outlined in Section \ref{Section3.1}. The larger the area under the curve, the more effective the parameter cut is at distinguishing between non-AGN transients and AGN. The dashed black line indicates the expected performance of a purely random classifier. The two-dimensional cut and the RMS cut are the most effective, whilst the host contrast parameter is the least effective.}
    \label{fig:ztf_roc}
\end{figure}

\begin{figure}
	\includegraphics[width=\columnwidth]{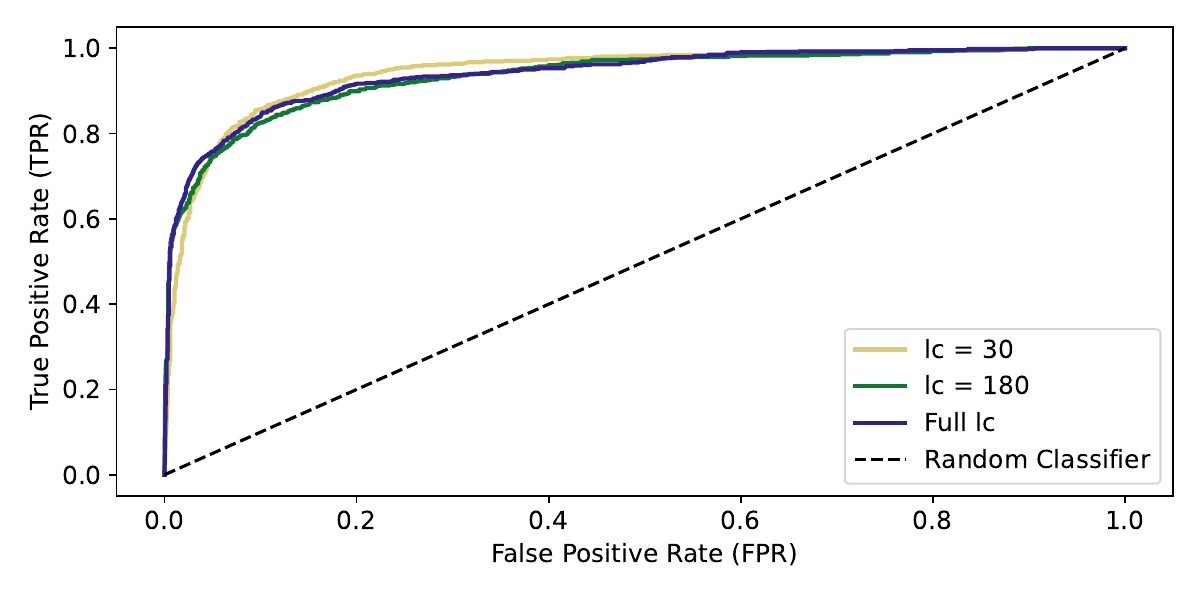}
    \caption{$g$ band ROC curves for the two-dimensional parameter cut described in Section \ref{Section3.1} calculated with the following amounts of light curve history: full light curve, 30\,d pre-detection and 180\,d pre-detection. The dashed black line indicates the expected performance of a purely random classifier. As is expected, the cut is most effective when the most history is available; however, it remains effective with lower amounts of light curve history.}
    \label{fig:ztf_time_roc}
\end{figure}

From the parameters tested above, the ratio of detection flux to pre-detection RMS is the most effective at distinguishing between AGN and other transients. This is likely because it is sensitive to both the mean flux of the pre-detection light curve and the scatter of the points in the light curve. 

However, for real-time implementation on LSST data, it is much more efficient and scalable to enact cuts based on summary information that is readily available in the alert packet, rather than computing new parameters from full light curves of every transient. Both the standard deviation and the mean flux of each band are present in the LSST alerts, whereas the RMS is not\footnote{See \hyperlink{https://sdm-schemas.lsst.io/apdb.html}{LSST Alert Schema}}. Therefore, whilst the RMS is the most effective individual cut, for implementation on real-time incoming LSST data, a combined cut using both the standard deviation and the mean flux that can achieve comparable effectiveness to the RMS is preferable.

The host contrast parameter is also effective at distinguishing AGN from other transients; however, it is dependent on knowing the innate host galaxy flux. This therefore limits its utility on real-time incoming LSST data as many transients will be from previously uncatalogued galaxies. However, it could be useful for studies on archival data.

We tested two-dimensional cuts based on ranges of thresholds for $F_{\rm det} / \sigma_F$ and $F_{\rm det} / \langle F \rangle$, respectively. For each possible combination, we calculated the resultant completeness, purity and F1 score values. The results of this test are shown in Figure \ref{fig:ztf_2d}. This two-dimensional cut using both the mean flux and the standard deviation, $F_{\rm det}/\sigma_F$-$F_{\rm det}/\langle F \rangle$, is able to achieve a peak F1 score of 0.78, with corresponding completeness and purity values of 0.75 and 0.81, respectively.

A comparison of the effectiveness of this combined cut and other previously tested cuts is shown as a receiver operating characteristic (ROC) curve in Figure \ref{fig:ztf_roc}. 
The two-dimensional cut incorporating pre-detection mean flux and pre-detection standard deviation produces the largest area under the curve in Figure \ref{fig:ztf_roc} and therefore is shown to be the most effective cut of those tested. Given its effectiveness and scalability using ratios only of parameters already present in LSST alert packets, we take this as the preferable approach for implementing on actual LSST data.

\subsection{Light Curve History Dependence}
\label{Section3.2}

\begin{figure*}
    \centering
	\includegraphics[width=\textwidth]{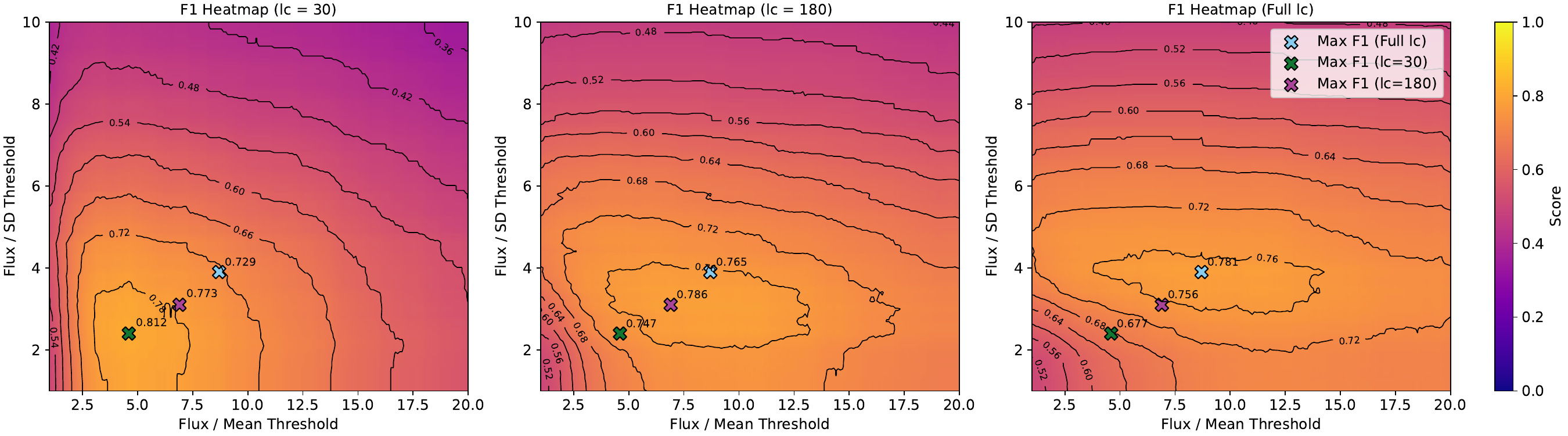}
    \caption{Heatmaps showing the F1 scores produced by a range of thresholds for the two-dimensional cut described in Section \ref{Section3.1} at different amounts of light curve history: 30\,d pre-detection (left), 180\,d pre-detection (centre), and the full light curve (right). The markers indicate the performance of the thresholds that maximise the F1 score for the 30\,d, 180\,d, and full-history light curves when applied to the data set shown in each panel. The cut which corresponds to the 180\,d pre-detection peak F1 score appears to be the most reliable cut, as it is effective at both small amounts of history and full light curve history. Whereas the 30\,d pre-detection cut is ineffective at full history, resulting from being too lenient and allowing too many AGN to pass, and the full history cut is ineffective at small amounts of history, as it is too strict, producing a low completeness.}
    \label{fig:ztf_heatmap_times}
\end{figure*}

The above calculations were conducted using the full pre-detection ZTF light curve for each object, often spanning multiple years. It is important to note that such data may not be available for every LSST transient: early in the survey or when the rolling cadence switches to new sky areas, it is unavoidable that there will be significantly less light curve history information available. Therefore, it is important to evaluate the impact of the availability of light curve history on the effectiveness of transient selection.

We compare the results using the full pre-detection ZTF light curve with those calculated using restricted light curves containing only 180 days or 30 days of pre-discovery forced photometry. Here, we evaluate the impact on the recommended $F/\sigma_F$-$F/\langle F \rangle$ cut preferable for LSST alerts, showing the ROC curves for the full and restricted light curves in Figure \ref{fig:ztf_time_roc}. For the light curve history dependence of the other parameter cuts, please see Appendix \ref{APP_lc_history}. 

Figure \ref{fig:ztf_time_roc} shows that at all values of light curve history the two-dimensional cut is broadly effective at distinguishing AGN from other transients. For the full light curve history, it is possible to achieve a higher true positive rate at a lower false positive rate, therefore achieving a higher purity. This therefore bolsters the utility of this parameter cut for real-time LSST data. It should be noted that there are different numbers of objects used to compute these ROC curves, as a small number of objects fail to have 30 days of pre-detection history and a larger number fail to have 180 days of pre-detection history. It is possible that this may impact the exact details of these ROC curves; however, the general trends should remain true. 

To evaluate the performance of the chosen parameter cut, $F_{\rm det}/\sigma_F$-$F_{\rm det}/\langle F \rangle$, in Figure \ref{fig:ztf_heatmap_times} we plot heatmaps showing the F1 scores produced by a range of thresholds for $F_{\rm det}/\sigma_F$ and $F_{\rm det}/\langle F \rangle$ - with each panel showing the results for a different amount of light curve history. The left-hand panel is constructed using 30 days of pre-detection history, the centre panel using 180 days, and the right-hand panel using the full pre-detection light curve history. For each set of light curve history we determine the optimal threshold values to maximise the F1 score. In Figure \ref{fig:ztf_heatmap_times}, we plot all three of these optimal threshold value positions on each panel with their corresponding F1 score stated to show the performance of these optimal threshold positions on differing amounts of light curve history.

Figure \ref{fig:ztf_heatmap_times} shows that the results produced by a respective threshold parameter change with the quantity of light curve history available. The shorter the duration of light curve history prior to detection, the lower the thresholds that maximise F1 score for both $F_{\rm det}/\sigma_F$ and$F_{\rm det}/\langle F \rangle$. This is especially evident when looking at the performance of the plotted optimal threshold positions on the panels showing other quantities of light curve history.

Although the highest total F1 score is produced by the optimal threshold for 30\,d of light curve history, the F1 score is inflated by high completeness (0.84) with the lenient threshold values chosen. This effect is most evident when the 30\,d optimal threshold values are applied to other quantities of light curve history, resulting in an F1 score of 0.75 when applied to 180 days of pre-detection light curve history and 0.68 when applied to the full light curve history.

It is unrealistic to expect full light curve history comparable to that of ZTF for LSST transients, given the gaps in the rolling cadence. 180 days is a more realistic expectation, and using the thresholds of $F_{\rm det}/\sigma_F$ and $F_{\rm det}/\langle F \rangle$ where the F1 score peaks in the central panel produces a reasonable F1 score across a wide range of pre-detection light curve history durations. Using the same 180\,d threshold values, $F_{\rm det}/\sigma_F > 3$ \& $F_{\rm det}/\langle F \rangle > 7$, produces an F1 score of 0.77 when applied to light curves with 30 days of history and 0.76 when applied to full light curves.

Therefore, we suggest implementing cuts using $F_{\rm det} / \sigma_F > 3$ and $F_{\rm det} / \langle F \rangle > 7$. For the full light curve history, these cuts produce an F1 score of 0.76, with completeness and purity values of 0.83 and 0.69, respectively. These chosen thresholds prioritise the ideal equally-weighted combination of both completeness and purity; however, if one of these is significantly more important than the other, these thresholds can be tuned to achieve the desired effect.

\subsection{Redshift Dependence}
\label{Section3.3}

The potential impact of redshift on the effectiveness of our chosen parameter cut was also evaluated. At larger redshift, and hence further distance from the observer, the general flux from a given object, and hence the signal-to-noise ratio of the data, may be diminished. This overall lower apparent brightness makes it more difficult to differentiate low-level but real physical variability from random noise. Therefore, it is important to test if our chosen parameter cuts remain effective on these objects.

\begin{figure}
	\includegraphics[width=\columnwidth]{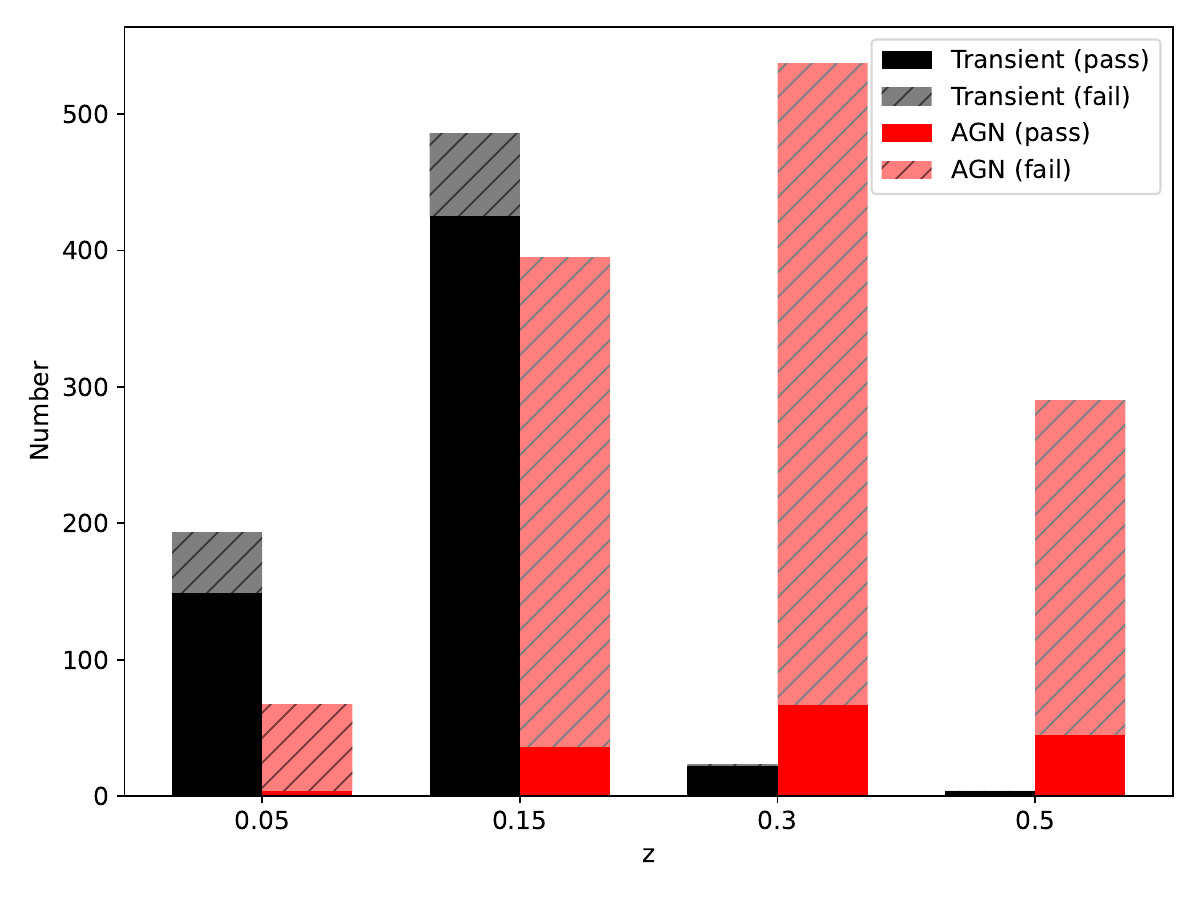}
    \caption{Stacked bar charts showing the performance of the two-dimensional cut described in Section \ref{Section3.1} on the labelled ZTF nuclear data set with respect to redshift. If a given object makes it through the filter, it is considered to have `passed' the cut. Whereas if it gets rejected by the cut, it is considered to have `failed'. In an ideal filter, all transients pass, and all AGN fail. The cut remains effective as redshift increases. The share of AGN with respect to other transients increases with the greater completeness of AGN samples at higher redshift.}
    \label{fig:redshift_ztf}
\end{figure}

\begin{table}
    \centering
    \caption{Results table outlining the performance of the two-dimensional cut on the ZTF labelled data set at different redshift ranges}
    \label{tab:ztf_redshift_table}
    \begin{tabular}{lll}
        \hline
        Redshift range & \% Transient pass & \% AGN pass \\
        \hline
        $z < 0.0.5$ & 77\% & 6\% \\
        $0.05 < z < 0.15$ & 87\% & 9\% \\
        $0.15 < z < 0.3$ & 96\% & 13\% \\
        $0.3 < z < 0.5$ & 100\% & 16\% \\
        \hline
    \end{tabular}
\end{table}

We separate the ZTF labelled sample into four groups: $z < 0.05$, $0.05 < z < 0.15$, $0.15 < z < 0.3$ and $0.3 < z < 0.5$. Boundaries for each bin are defined to ensure statistically useful numbers of objects in each bin. As demonstrated in Figure \ref{fig:redshift_ztf}, the chosen parameter cut is effective at selecting transients and rejecting AGN in all the redshift bins. The percentages of transients and AGN which successfully pass the two-dimensional cut at a given redshift range are outlined in Table \ref{tab:ztf_redshift_table}. While the chosen parameter cut remains effective at selecting transients, the share of contaminant AGN which pass does appear to increase slightly with increasing redshift. However, there are a greater number of AGN relative to transients as the redshift increases. This is due to a combination of the intrinsic luminosity of AGN, which is typically larger than most transients and enables detection to larger distances, and the completeness of the \texttt{Milliquas} catalogue, which is better than that for transient surveys in higher redshift bins (a Type Ia supernova at $z>0.15$ peaks at $\gtrsim19.5$\,mag, and is therefore below the limiting magnitude of the ZTF Bright Transient Survey \citep[BTS;][]{Perley2020}. 

At these larger redshift bins, our method still successfully passes transients whilst rejecting the majority of AGN. But we note that the fraction of AGN contaminants is likely to decrease for deeper surveys that probe higher redshifts. We will examine this further in Section \ref{Section5}.

\subsection{ZTF Unlabelled Sample Testing}

To emulate the implementation of this photometric variability parameter cut approach on the real-time flow of unknown transients, we tested the performance of the chosen cut on a subset of the ZTF blind sample -- i.e. those with no labels from the \texttt{TNS} or \texttt{Milliquas}. 

A manual process of visual inspection was used on a randomly selected sample of 3,000 unlabelled light curves from the blind sample. Upon inspection, each object was defined as either a high-amplitude transient that a typical observer would manually select for spectroscopic follow-up, or a visibly stochastic light curve that would not be manually selected for spectroscopic follow-up. The former and latter serve as proxies for non-AGN transients and AGN, respectively. While this process is of course imperfect (we cannot say for certain which unlabelled objects are indeed transients), it closely resembles the process by which transients have been selected for follow-up historically, prior to magnitude-limited surveys like the ZTF Bright Transient Survey or TiDES.

The effectiveness of four different parameter cuts, outlined in sections \ref{Section2.3}, \ref{Section2.4}, \ref{Section2.5} \& \ref{Section3.1} was evaluated on this pseudo-labelled subset of the blind ZTF nuclear data set: $F_{\rm det} / \sigma_F$, $F_{\rm det} / \langle F \rangle$, $F_{\rm det} / RMS_F$, and our chosen two-dimensional parameter cut $F_{\rm det}/\sigma_F$-$F_{\rm det}/\langle F \rangle$. The threshold values chosen and resultant F1 scores, completeness values and purity values are shown in Table \ref{tab:blind_results_table}.

\begin{table}
    \centering
    \caption{Results table showing effectiveness of each parameter applied to the blind ZTF nuclear sample.}
    \label{tab:blind_results_table}
    \begin{tabular}{lllll}
        \hline
        Parameter & Threshold & F1 & Completeness & Purity \\
        \hline
        $F_{\rm det} / \sigma_F$ & 3 & 0.73 & 0.87 & 0.63 \\
        $F_{\rm det} / \langle F \rangle$ & 7 & 0.72 & 0.93 & 0.58 \\
        $F_{\rm det} / RMS_F$ & 4 & 0.77 & 0.71 & 0.84 \\
        $F_{\rm det}/\sigma_F$-$F_{\rm det}/\langle F \rangle$ & 3 - 7 & 0.79 & 0.85 & 0.74 \\
        \hline
    \end{tabular}
\end{table}

Similar to the results shown in Figure \ref{fig:ztf_roc}, the two-dimensional cut, $F_{\rm det}/\sigma_F$-$F_{\rm det}/\langle F \rangle$, is the most effective, achieving a peak F1 score of 0.79 with completeness and purity values of 0.85 and 0.74, respectively. $F_{\rm det} / RMS_F$ achieves similar success, with a peak F1 score of 0.77 produced by a greater purity value (0.84) but lower completeness (0.71). $F_{\rm det} / \sigma_F$ and $F_{\rm det} / \langle F \rangle$ assessed individually achieve high completeness scores but low purity, resulting in F1 scores of 0.73 and 0.72, respectively. These results indicate that these parameter cuts remain effective at distinguishing high-amplitude transients from stochastic variation when applied to unlabelled data, automating a process that has historically been done by eye in smaller surveys, but will not be possible at the scale of LSST.

\section{MALLORN Data Testing}
\label{Section4}
\subsection{Effectiveness on Simulated LSST Data}

\begin{figure*}
    \centering
	\includegraphics[width=\textwidth]{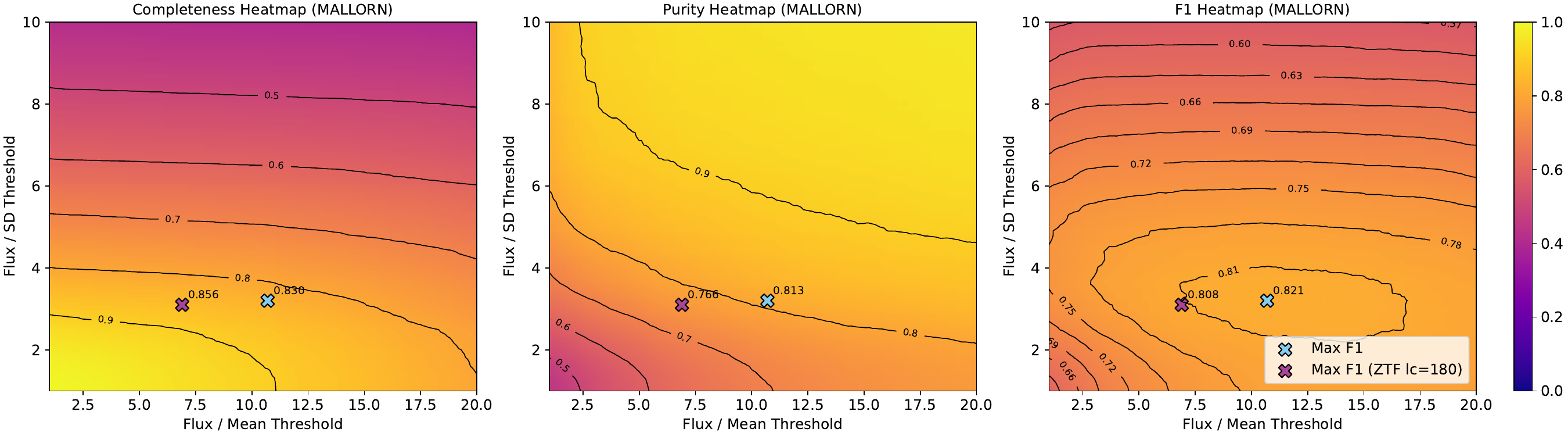}
    \caption{Heatmaps indicating the completeness (left), purity (centre) and F1 score produced by a range of cuts on the MALLORN data set for the ratio between detection flux and pre-detection standard deviation (y-axis) and the ratio between detection flux and pre-detection mean flux (x-axis). The peak F1 score value (indicated by the blue marker) occurs at a value of 0.82, with corresponding completeness and purity values of 0.83 and 0.81, respectively. The performance of the chosen cut parameters based on the optimal 180\,d thresholds is indicated with the purple marker. It achieves a comparable F1 score of 0.81, with corresponding completeness and purity values of 0.86 and 0.77, respectively.}
    \label{fig:mallorn_heatmap}
\end{figure*}

\begin{figure}
	\includegraphics[width=\columnwidth]{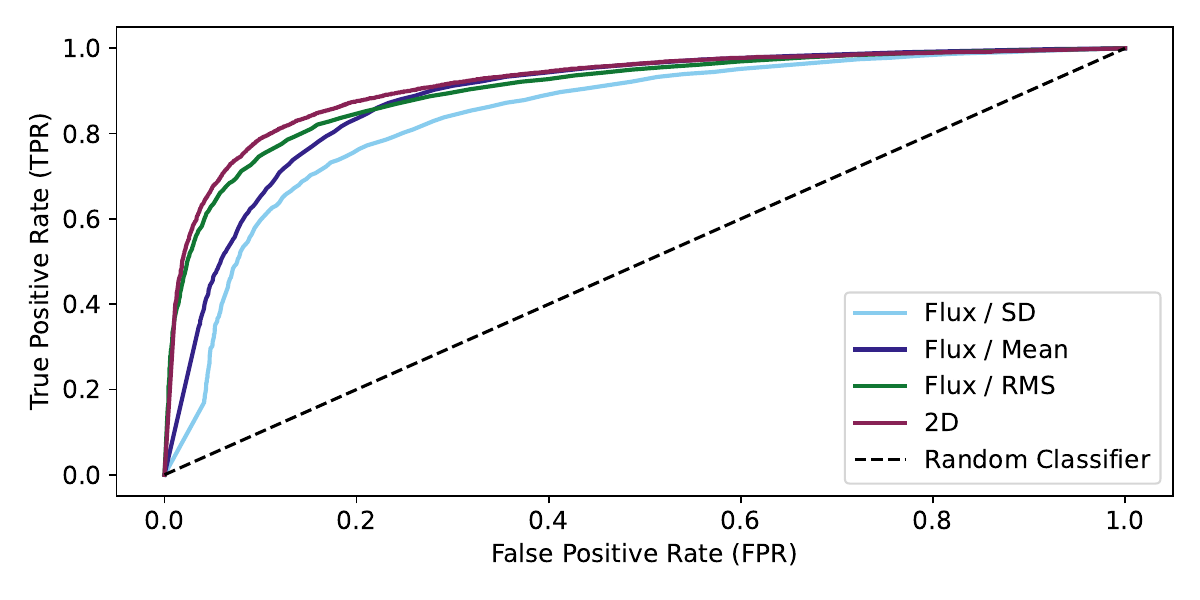}
    \caption{ROC curves plotting the true positive rate against the false positive rate for the following parameter cuts applied to the MALLORN data set: $F_{\rm det} / \sigma_F$, $F_{\rm det} / \langle F \rangle$, $F_{\rm det} / RMS_F$, and the two-dimensional cut, $F_{\rm det}/\sigma_F$-$F_{\rm det}/\langle F \rangle$, outlined in Section \ref{Section3.1}. The dashed black line indicates the expected performance of a purely random classifier. $F_{\rm det}/\sigma_F$-$F_{\rm det}/\langle F \rangle$ and $F_{\rm det} / RMS_F$ are the most effective, whilst $F_{\rm det} / \sigma_F$ is the least effective.}
    \label{fig:mallorn_ROC}
\end{figure}

Thus far, we have only evaluated the effectiveness of the chosen photometric variability parameters on ZTF data. To validate that this approach is similarly effective for LSST data, we conduct the same series of tests described in Section \ref{Section3} on the simulated LSST data set MALLORN \citep[][]{MALLORN}. 

There are several differences between the photometry from ZTF and LSST. The Vera C. Rubin Observatory, conducting LSST, has a much larger mirror (8.4m) than the Palomar 48-inch Schmidt Telescope \citep[][]{Harrington1952} used for ZTF, and consequently LSST can see to much greater depth and probe fainter sources not accessible to ZTF. Therefore, LSST probes a larger volume, and within a given volume will produce many transient detections at lower luminosities. As our selection strategies use the \emph{ratios} between detection flux and the variability parameters, we expect that their effectiveness should be invariant to the distance to the sources -- provided of course that the more distant source remains above the survey detection limit. Through testing on the MALLORN data set we can confirm this. 

Another key difference between ZTF and LSST photometry is in the filters used. ZTF observed primarily in $g$ and $r$ (with some observations in $i$, though insufficient in number to be included in our ZTF nuclear data set), whereas LSST conducts observations in the $u$, $g$, $r$, $i$, $z$ \& $y$ bands. The LSST Wide Fast Deep (WFD) observes a target field with two filters during a night, with at least one of those filters being $g$, $r$ or $i$ band. The survey will, conditions permitting, revisit that location with at least one of the same filters three nights later \citep[][]{LSST,Jones2021,Bianco2022}. 

There are distinct differences in the depths of each band, with expected single-exposure detection limits of $m_u = 23.8$, $m_g = 24.5$, $m_r = 24.0$, $m_i = 23.4$, $m_z = 22.7$ \& $m_y = 22.0$ \citep[][]{LSSTCam}. Consequently, particularly for observations of fainter transients, some bands are more useful for transient selection than others. Photometric noise in a shallow filter may artificially increase the apparent RMS variability, for example. 

Based on the trade-off between cadence and signal-to-noise, we must decide which combination of filters to use when determining the photometric variability parameters. We find that the optimal combination is to use the $g$, $r$ \& $i$ bands, since observations in these bands achieve the greatest (and comparable) depth, have a relatively uniform cadence when combined, and have relatively small spread in wavelength to minimise the impact of degeneracies between colour and variability. For analysis on the utility of different filter combinations, please see Appendix \ref{APP_filter_choice}.

To combine the data from separate bands, we effectively treat $g$, $r$ and $i$ as a single band, and compute the mean and standard deviation of all fluxes measured in $g$, $r$ or $i$. When applying this to real-time LSST alerts, one can simply compute a weighted mean and combined standard deviation from the statistics provided in the alert packets for each of the $g$, $r$ and $i$ band light curves.

Applying the same approach as in Section \ref{Section3.1}, we produce heatmaps of the completeness, purity and F1 score produced by our recommended two-dimensional cut, $F_{\rm det}/\sigma_F$-$F_{\rm det}/\langle F \rangle$, applied to the MALLORN data. These are shown in Figure \ref{fig:mallorn_heatmap}. Implementing the suggested thresholds based on the optimal ZTF 180\,d cut and applying this to the MALLORN data produces an F1 score of 0.81, with corresponding completeness and purity values of 0.86 and 0.77, respectively. These results indicate that this approach utilising photometric variability parameters to distinguish non-AGN from AGN remains effective when applied to LSST-like data.

An ROC curve comparing the effectiveness of the two-dimensional cut compared to individual photometric variability parameters is shown in Figure \ref{fig:mallorn_ROC}. As the MALLORN data set does not contain simulated host luminosities, it was not possible to evaluate the performance of the host contrast parameter, $\Delta m_{\mathrm{var}}$, on the MALLORN data. Similarly to Figure \ref{fig:ztf_roc}, $F_{\rm det}/\sigma_F$-$F_{\rm det}/\langle F \rangle$ and $F_{\rm det} / RMS_F$ are the most effective, whilst the $F_{\rm det} / \sigma_F$ is the least effective.

\subsection{Light Curve History Dependence}

\begin{figure}
	\includegraphics[width=\columnwidth]{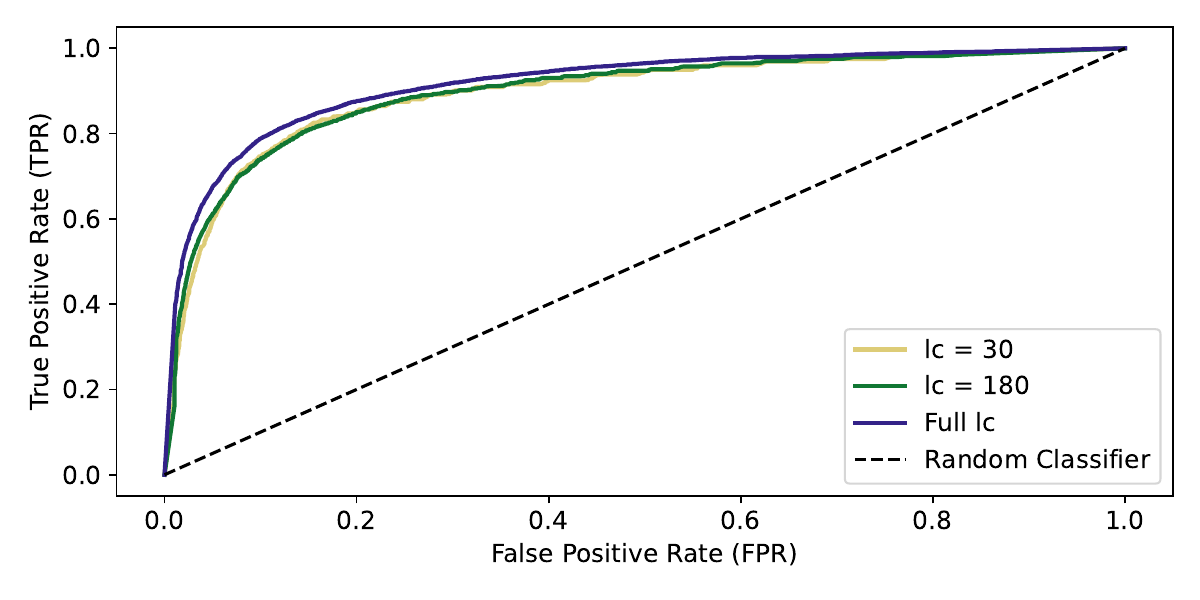}
    \caption{ROC curves for the two-dimensional parameter cut described in Section \ref{Section3.1} calculated with the following amounts of MALLORN light curve history: full light curve, 30\,d pre-detection and 180\,d pre-detection. The dashed black line indicates the expected performance of a purely random classifier. As is expected, the cut is most effective when the most history is available; however, it remains effective with lower amounts of light curve history.}
    \label{fig:mallorn_ROC_time}
\end{figure}

To evaluate the impact of the amount of light curve history available for LSST data, we replicate the testing in Section \ref{Section3.2} on the MALLORN data. Figure \ref{fig:mallorn_ROC_time} shows the same trends found in \ref{fig:ztf_time_roc}. The two-dimensional cut is most effective when the most history is available; however, it remains effective with lower amounts of light curve history. This suggests that even early in LSST operations, when we have comparatively little light curve history, implementing this approach to distinguish non-AGN transients from standard AGN variability should remain effective.

It is possible that the inclusion of AGN light within template images may have an impact on the effectiveness of some light curve parameters. The $F_{\rm det}/\sigma_F$ portion of the cut will likely be largely unaffected, as any subsequent difference imaging as any AGN light in the template will only shift the baseline flux level in the difference images, but flux measurements will still show a larger standard deviation than a quiescent galaxy. This is only likely to be a concern in the time immediately following the template image (\textasciitilde days-weeks), in which variability of the AGN from its level in the template may be modest. The existence of AGN light within the template is of more concern for the $F_{\rm det}/\langle F \rangle$ portion of the cut. As the absolute value of the mean baseline flux is used for calculating the ratio, the presence of a baseline offset could reduce the apparent contrast. However, such a large shift in baseline (i.e.~comparable to the brightness of a genuine transient) is only likely to occur in very bright and highly variable AGN, and such cases are less likely be genuine transients -- and unlikely to pass the standard deviation cut regardless. Therefore, whilst this is a concern to be aware of, it is unlikely to have a major impact on our results. We note that this effect is also present in the ZTF data (since it contains AGN) and therefore is included in our existing evaluations.

\subsection{Redshift Dependence}

\begin{figure}
	\includegraphics[width=\columnwidth]{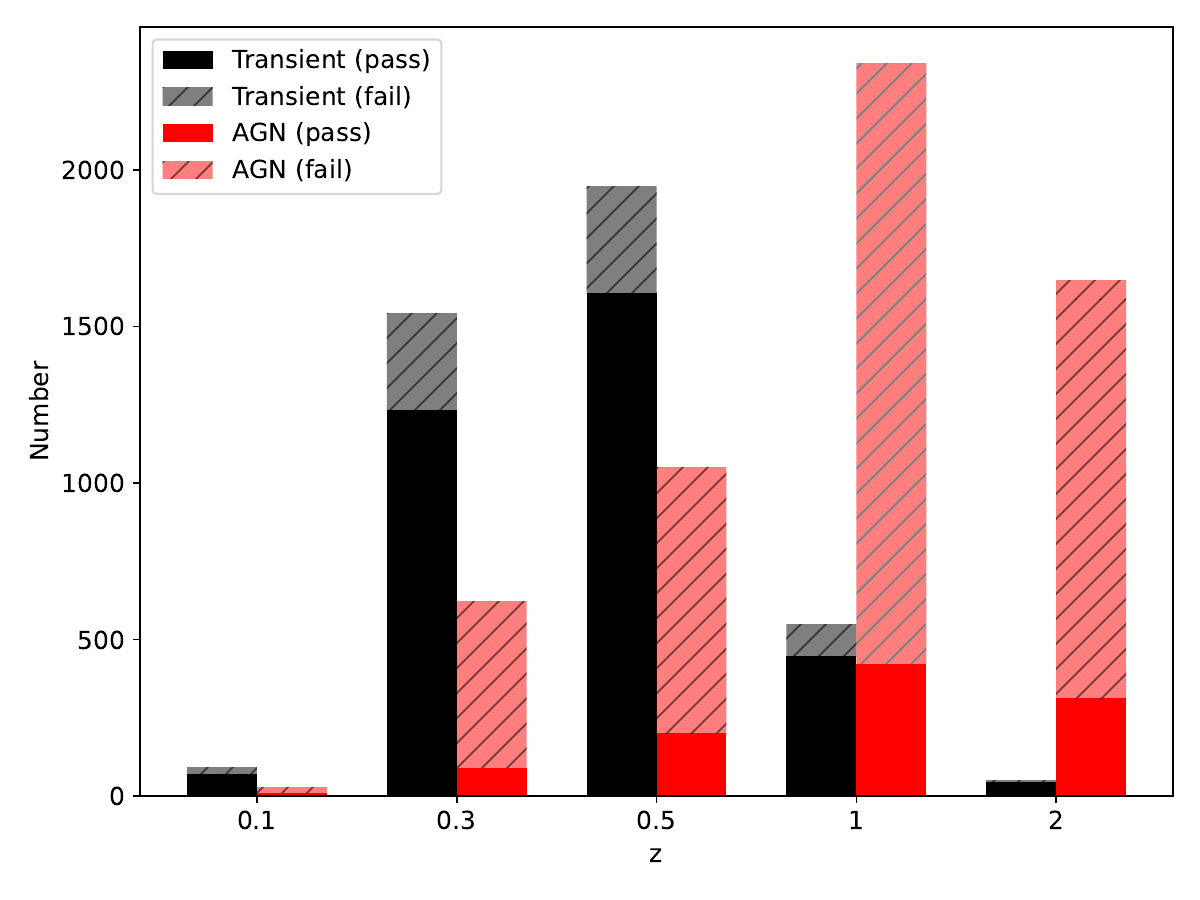}
    \caption{Stacked bar charts showing the performance of the two-dimensional cut described in Section \ref{Section3.1} on the MALLORN data set with respect to redshift. The cut remains effective as redshift increases. The share of AGN increases with redshift due to the presence of AGN identified via \texttt{Milliquas} in the light curves used to produce MALLORN.}
    \label{fig:redshift_mallorn}
\end{figure}

We emulate the redshift analysis conducted in Section \ref{Section3.3} for the MALLORN data. As the MALLORN data is simulating LSST transients, it contains objects at a significantly greater redshift than those within the ZTF nuclear sample. Consequently, we separate the MALLORN data set into five groups: $z < 0.1$, $0.1 < z < 0.3$, $0.3 < z < 0.5$, $0.5 < z < 1$ and $1 < z < 2$. The results of this analysis are shown in Figure \ref{fig:redshift_mallorn} and summarised in Table \ref{tab:mallorn_redshift_table}. The overall trend is similar to that of \ref{fig:redshift_ztf}, with the chosen two-dimensional parameter cut largely effective in all redshift bins. By construction, there are a greater number of simulated AGN relative to transients as the redshift increases, as the MALLORN data set is constructed from the ZTF nuclear data set. The two-dimensional cut remains largely effective at passing transients and excluding the majority of AGN at these higher redshift bins, with only a small subset of the AGN passing the cut.

\begin{table}
    \centering
    \caption{Results table outlining the performance of the two-dimensional cut on the MALLORN data set at different redshift ranges}
    \label{tab:mallorn_redshift_table}
    \begin{tabular}{lll}
        \hline
        Redshift range & \% Transient pass & \% AGN pass \\
        \hline
        $z < 0.1$ & 79\% & 37\% \\
        $0.1 < z < 0.3$ & 80\% & 15\% \\
        $0.3 < z < 0.5$ & 82\% & 19\% \\
        $0.5 < z < 1$ & 82\% & 18\% \\
        $1 < z < 2$ & 88\% & 19\% \\
        \hline
    \end{tabular}
\end{table}

\section{Discussions \& Conclusions}
\label{Section5}
The two-dimensional cut utilising both $F_{\rm det}/\sigma_F$ and $F_{\rm det}/\langle F \rangle$ is a demonstrably effective means to distinguish non-AGN transients from standard AGN variability. Using $F_{\rm det} / \sigma_F > 3$ and $F_{\rm det} / \langle F \rangle > 7$ (based on the optimal thresholds for 180\,d of light curve history in ZTF) has been proven to be effective on three separate data sets. These results are summarised in Table \ref{tab:summary_table}. Producing an F1 score of 0.76 (completeness = 0.83, purity = 0.69) on the labelled ZTF nuclear data set demonstrates that the two-dimensional cut is reliably effective at separating a variety of non-AGN transients from standard AGN variability. 
The effective performance on the blind ZTF nuclear data set (F1 = 0.79, completeness = 0.85, purity = 0.74) validates the effectiveness of the two-dimensional cut on unclassified data to select events that would be manually chosen for spectroscopy.
The success when tested on the MALLORN data set (F1 = 0.81, completeness = 0.86, purity = 0.77) indicates that the two-dimensional cut will be effective on LSST-like photometry. This verifies that this approach remains effective even with less frequent observing cadence and when applied to a selection of multiple wavelength bands. The most effective set of wavelength bands was found to be $g$, $r$ \& $i$. It is likely that supplementary photometry from the La Silla Schmidt Southern Survey \citep[LS4;][]{LS4}, enhancing the coverage of the $g$ and $i$ bands, will further assist the identification of variability within LSST detections. 
Further testing verified that this approach remains effective at selecting transients and rejecting the majority of AGN at redshift $z>0.5$ and with varying amounts of light curve history.

Based on the above results, we expect that implementing this approach to avoid AGN contamination in LSST transient target selection would produce a sample with a completeness and a purity both $\gtrsim0.7$-0.8. The thresholds used on $F_{\rm det}/\sigma_F$ and $F_{\rm det}/\langle F \rangle$ can be adjusted to further prioritise either completeness or purity as desired; however, it is important to note that enhancing one will detract from the other.

Using TiDES as an example, we can justify the need for implementing a means of limiting AGN contamination as described in this paper. AGN have an approximate sky density of 344 AGN/deg$^2$ to an $r$ band magnitude limit of 23.5 \citep[][]{DeCicco2021}. This rate can be rescaled to that of the volume produced by the TiDES detection limit (22.5) producing a value of 86.4 AGN/deg$^2$. 


 
 There are on average \textasciitilde5-10 SNe in the 4.2 deg$^2$ 4MOST field-of-view. Based on the above calculation, we can expect \textasciitilde350 AGN in a 4MOST pointing. Therefore, there are approximately 35x more AGN than SNe observable by TiDES in a given pointing. Without the implementation of a means to distinguish alerts due to AGN variability from those of transients of interest, it is likely that AGN contamination would significantly diminish the potential scientific output of TiDES.


\begin{table}
    \centering
    \caption{Summary results table showing performance of chosen two-dimensional parameter, $F_{\rm det}/\sigma_F$-$F_{\rm det}/\langle F \rangle$, on the ZTF labelled, ZTF blind and MALLORN data sets}
    \label{tab:summary_table}
    \begin{tabular}{llll}
        \hline
        Data & F1 & Completeness & Purity \\
        \hline
        ZTF (labelled) & 0.76 & 0.83 & 0.69 \\
        ZTF (blind) & 0.79 & 0.85 & 0.74 \\
        MALLORN & 0.81 & 0.86 & 0.77 \\
        \hline
    \end{tabular}
\end{table}

We can conclusively state that it is possible to use simple variability parameters to distinguish between detections of stochastically varying standard AGN activity and detections of transients of interest (such as SNe and TDEs). This approach is particularly useful for statistical surveys when it is necessary to be able to accurately model the selection function. Additionally, this could be included as a feature to distinguish AGN from non-AGN in a photometric light curve classifier, whilst the inverse of the chosen approach could be used to photometrically select for AGN.

\section*{Acknowledgements}

Thanks to M. Kowalski and S. Reusch for running the \texttt{AMPEL} Nuclear Filter and creating the ZTF nuclear transient data set used to create the MALLORN data set. Thanks to M. Schwamb for the suggestion to investigate the impact of light curve duration.
DM acknowledges a studentship funded by the Leverhulme Interdisciplinary Network on Algorithmic Solutions.
DM acknowledges travel support provided by STFC for UK participation in LSST through grant ST/X001334/1.
MN and CRA are supported by the European Research Council (ERC) under the European Union’s Horizon 2020 research and innovation programme (grant agreement No.~948381). 
PW is grateful for the support from the Science and Technology Facilities Council (STFC) grant ST/Z510269/1.

\section*{Data Availability}

The ZTF photometry used within this paper is publicly available from the ZTF Forced Photometry Server. The MALLORN data set is publicly accessible via Kaggle (\url{https://www.kaggle.com/competitions/mallorn-astronomical-classification-challenge/overview}). The code relevant to this paper has been uploaded to a GitHub repository (\url{https://github.com/dkjmagill/AGN_Phot_Var_Screening}) and an implementation of the $F_{\rm det}/\sigma_F$-$F_{\rm det}/\langle F \rangle$ two-dimensional cut will be made available as a public filter on Lasair.



\bibliographystyle{mnras}
\bibliography{tides_var} 




\appendix

\section{WISE AGN Separation}
\label{APP_WISE}
\begin{figure}
	\includegraphics[width=\columnwidth]{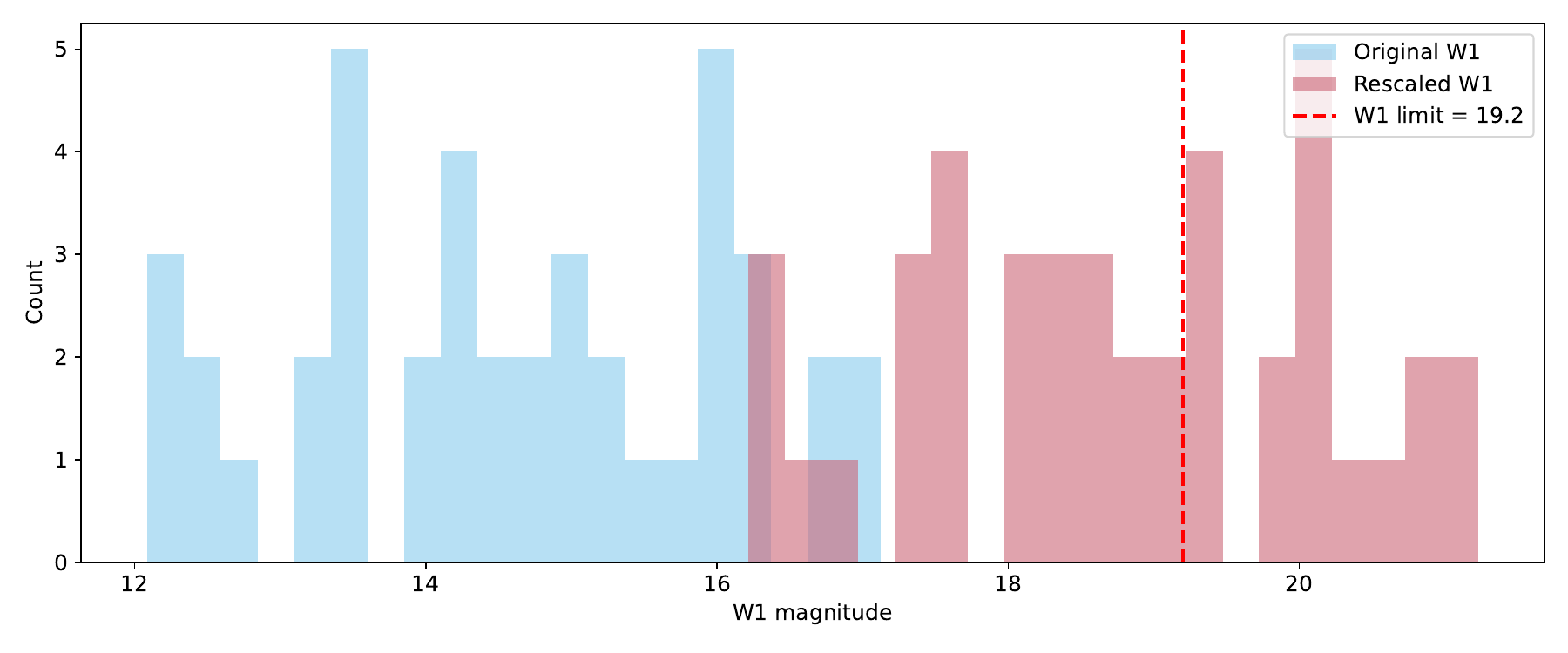}
    \caption{Plot of $W1$ magnitudes for all TDE host galaxies in the ZTF nuclear sample and their $W1$ magnitudes when rescaled to the redshifts of the corresponding simulated TDEs in MALLORN. The dashed red line indicates the $W1$ limiting magnitude of \textit{WISE}. A large number of rescaled $W1$ magnitudes are beyond the \textit{WISE} detection limit. Therefore, indicating that the $W1$ - $W2$ colour criterion likely will not be useful for a lot of LSST discoveries.}
    \label{fig:wise_analysis}
\end{figure}

A mid-infrared colour criterion using \textit{Wide-field Infrared Survey Explorer} \citep[\textit{WISE;}][]{WISE} bands $W1$ ($3.4\mu m$) - $W2$ ($4.6\mu m$) $>$ 0.8 has previously been used to distinguish AGN from non-AGN. We evaluated the potential of using this approach on LSST data. We selected all of the TDEs within the ZTF nuclear sample which produced a simulated MALLORN light curve which peaked at an $r$ band magnitude above 22.5 and queried the AllWISE catalogue \citep[][]{AllWISE} to retrieve the $W1$ and $W2$ band magnitudes of their host galaxies. We then rescale these magnitudes according to the redshift values of the simulated TDEs in MALLORN, using the approach outlined in Section 2.3 of \citet{MALLORN}. The original $W1$ values and the rescaled $W1$ values for LSST objects are shown in Figure \ref{fig:wise_analysis}.

A large number of rescaled $W1$ magnitudes (\textasciitilde30-50\%) are beyond the \textit{WISE} AB detection limit (19.2). This therefore suggests that the $W1$ - $W2$ colour criterion will not be useful for distinguishing AGN from non-AGN for a large number of LSST detections, as the \textit{WISE} magnitudes of their host galaxies will be fainter than the \textit{WISE} detection limits. In order for this previously highly successful approach to be extended to implementation on LSST discoveries, a new \textit{WISE}-style telescope with magnitude limits of greater depth would be required.

\section{ZTF $r$ band results}
\label{APP_ZTF_r}
\begin{figure}
	\includegraphics[width=\columnwidth]{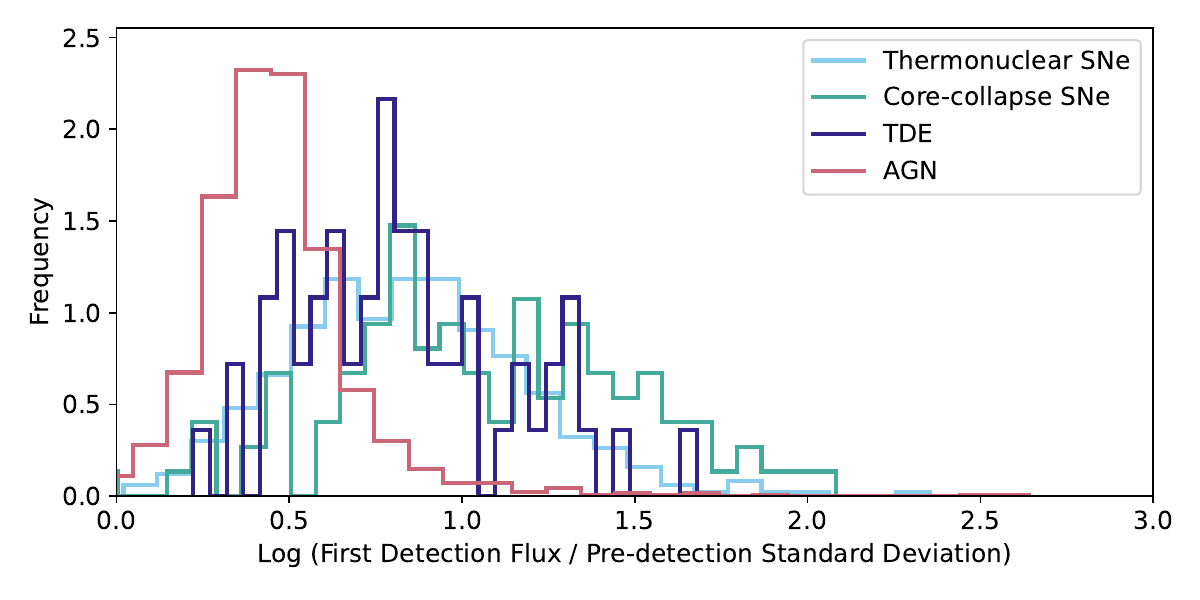}
    \caption{Histogram showing the $r$ band distribution of the log ratio of first detection flux to pre-detection standard deviation for thermonuclear supernovae, core-collapse supernovae, tidal disruption events and active galactic nuclei. There is a clear separation between the peak of the AGN distribution (\textasciitilde 0.3) and that of the other transient types (\textasciitilde 0.8). This demonstrates that this parameter is also effective in the $r$ band as a metric to distinguish AGN from other transients.}
    \label{fig:r_std_hist}
\end{figure}

\begin{figure}
	\includegraphics[width=\columnwidth]{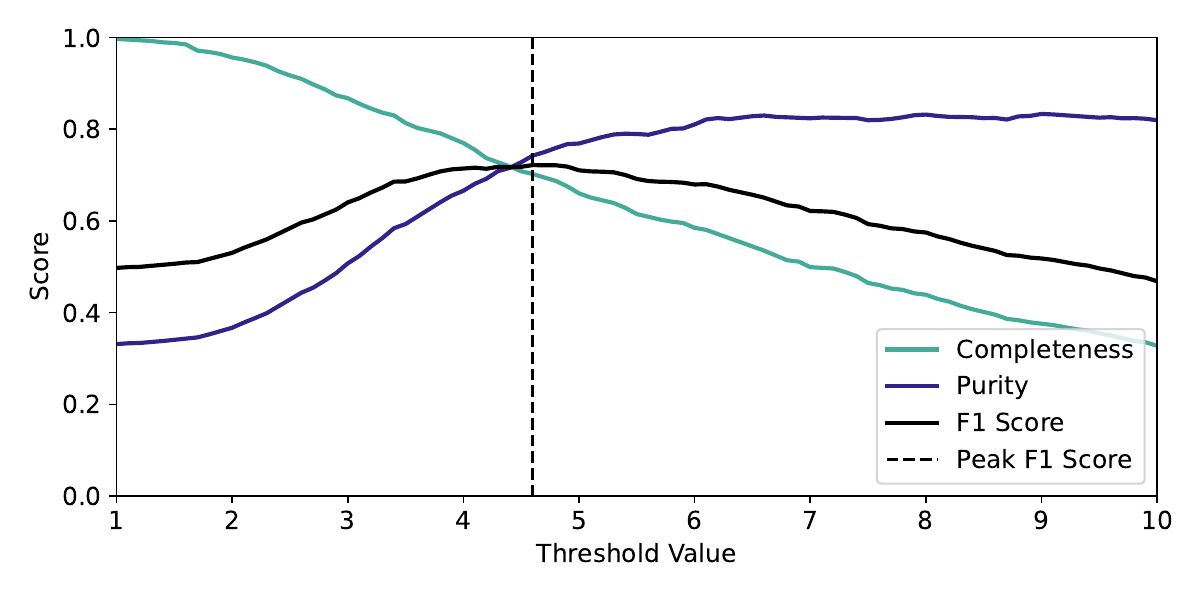}
    \caption{Plot of completeness, purity and F1 score against parameter threshold value for the $r$ band ratio of detection flux to pre-detection flux standard deviation. The F1 score peaks at a value of 0.72 at threshold value 4.6 (indicated by the dashed line), with corresponding completeness and purity values of 0.70 and 0.74, respectively.}
    \label{fig:r_std_cpf1}
\end{figure}

\begin{figure}
	\includegraphics[width=\columnwidth]{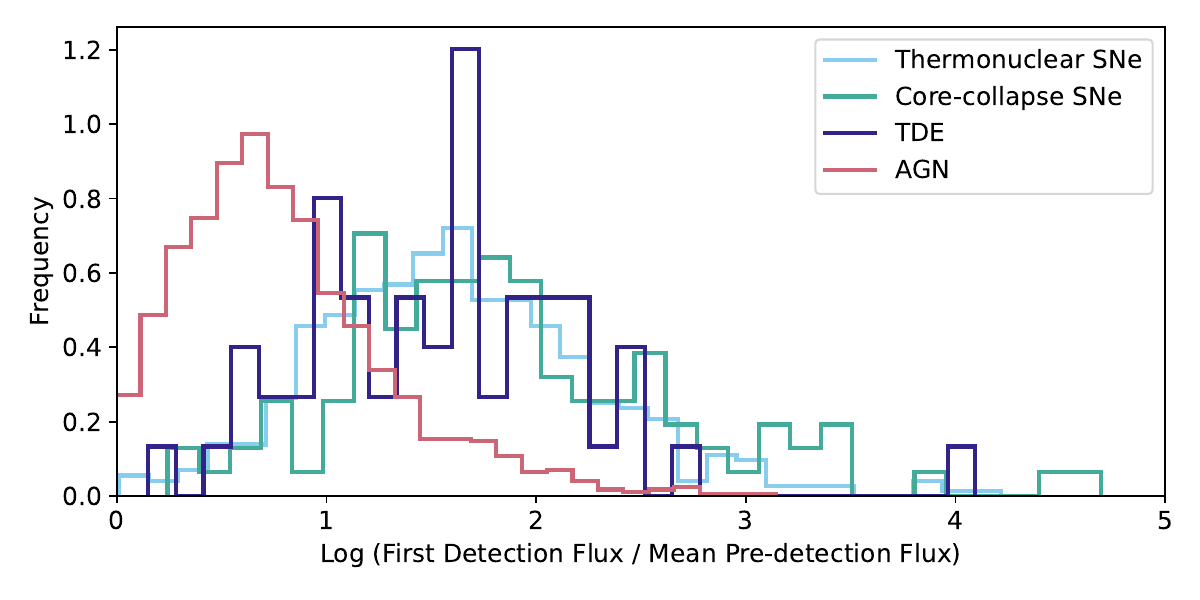}
    \caption{Histogram showing the $r$ band distribution of the log ratio of first detection flux to pre-detection mean flux for thermonuclear supernovae, core-collapse supernovae, tidal disruption events and active galactic nuclei. There is a clear separation between the peak of the AGN distribution (\textasciitilde 0.7) and that of the other transient types (\textasciitilde 1.7). This demonstrates that this parameter is also effective in the $r$ band as a metric to distinguish AGN from other transients.}
    \label{fig:r_mf_hist}
\end{figure}

\begin{figure}
	\includegraphics[width=\columnwidth]{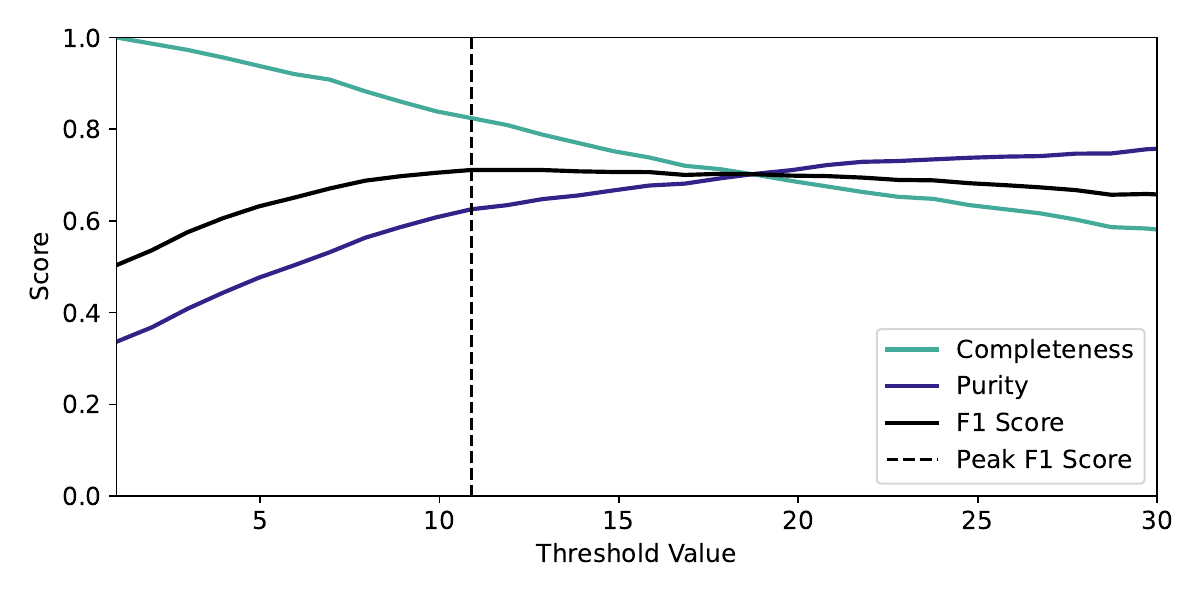}
    \caption{Plot of completeness, purity and F1 score against parameter threshold value for the $r$ band ratio of detection flux to pre-detection flux mean. The F1 score peaks at a value of 0.71 at threshold value 10.9 (indicated by the dashed line), with corresponding completeness and purity values of 0.82 and 0.63, respectively.}
    \label{fig:r_mf_cpf1}
\end{figure}

\begin{figure}
	\includegraphics[width=\columnwidth]{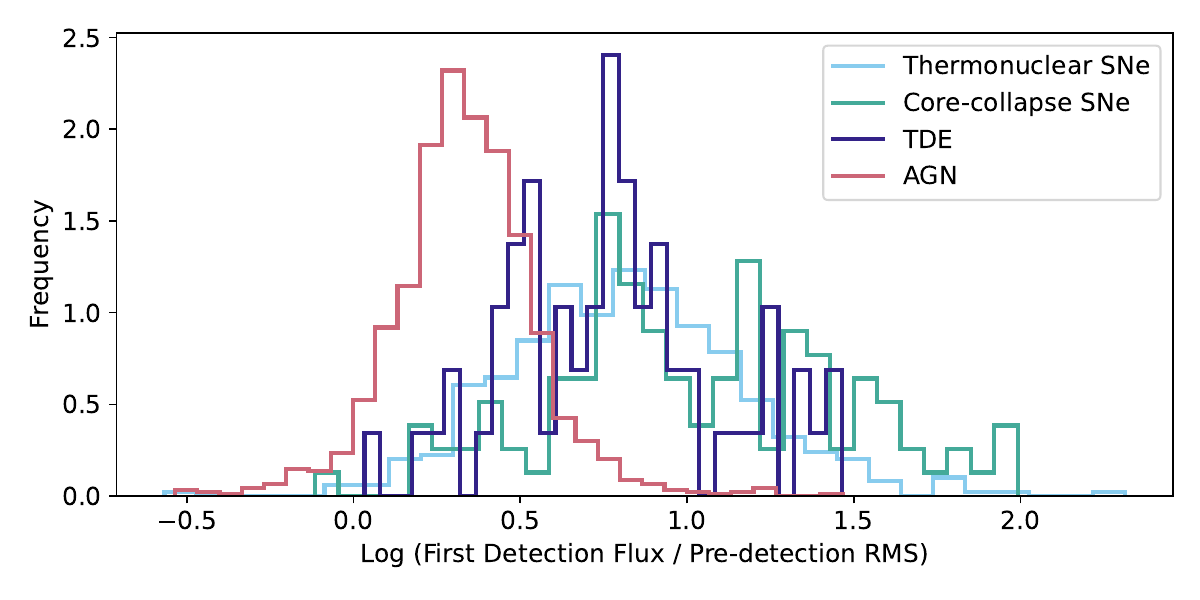}
    \caption{Histogram showing the $r$ band distribution of the log ratio of first detection flux to pre-detection root mean square for thermonuclear supernovae, core-collapse supernovae, tidal disruption events and active galactic nuclei. There is a clear separation between the peak of the AGN distribution (\textasciitilde 0.3) and that of the other transient types (\textasciitilde 0.8). This demonstrates that this parameter is also effective in the $r$ band as a metric to distinguish AGN from other transients.}
    \label{fig:r_rms_hist}
\end{figure}

\begin{figure}
	\includegraphics[width=\columnwidth]{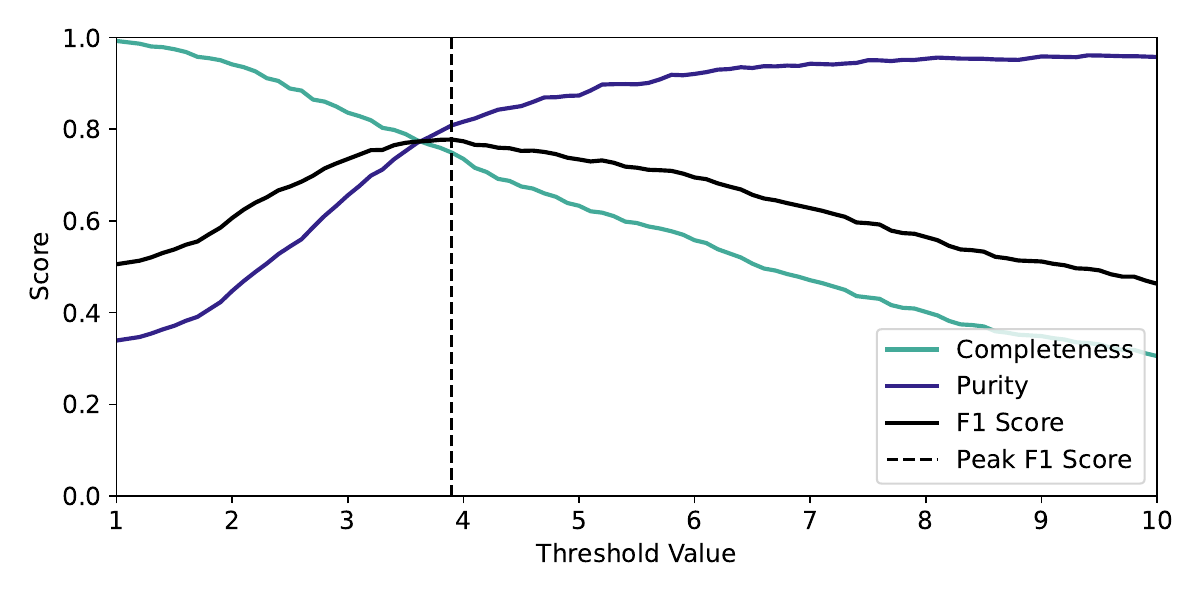}
    \caption{Plot of completeness, purity and F1 score against parameter threshold value for the $r$ band ratio of detection flux to pre-detection flux RMS. The F1 score peaks at a value of 0.78 at threshold value 3.9 (indicated by the dashed line), with corresponding completeness and purity values of 0.75 and 0.81, respectively.}
    \label{fig:r_rms_cpf1}
\end{figure}

\begin{figure}
	\includegraphics[width=\columnwidth]{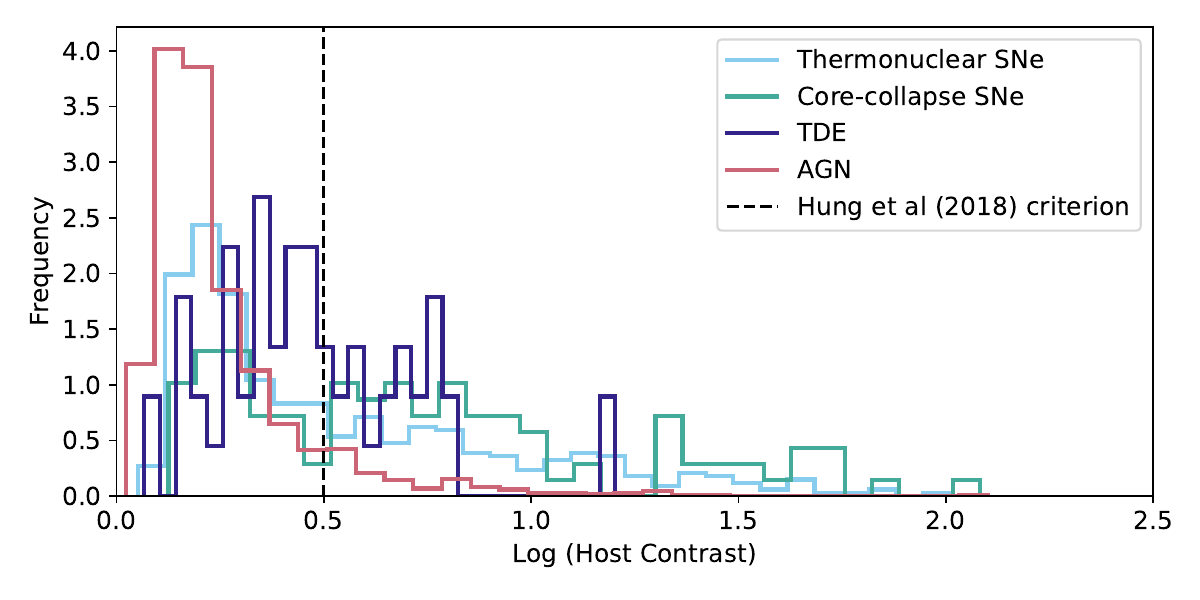}
    \caption{Histogram showing the $r$ band distribution of the log of the host contrast parameter defined in \citet{Hung2018} for thermonuclear supernovae, core-collapse supernovae, tidal disruption events and active galactic nuclei. There is a clear separation between the peak of the AGN distribution (\textasciitilde 0.2) and that of the other transient types (\textasciitilde 0.5). The threshold value of 0.5 implemented in \citet{Hung2018} is indicated with a dashed black line. This demonstrates that this parameter is also effective in the $r$ band as a metric to distinguish AGN from other transients.}
    \label{fig:r_host_contrast_hist}
\end{figure}

\begin{figure}
	\includegraphics[width=\columnwidth]{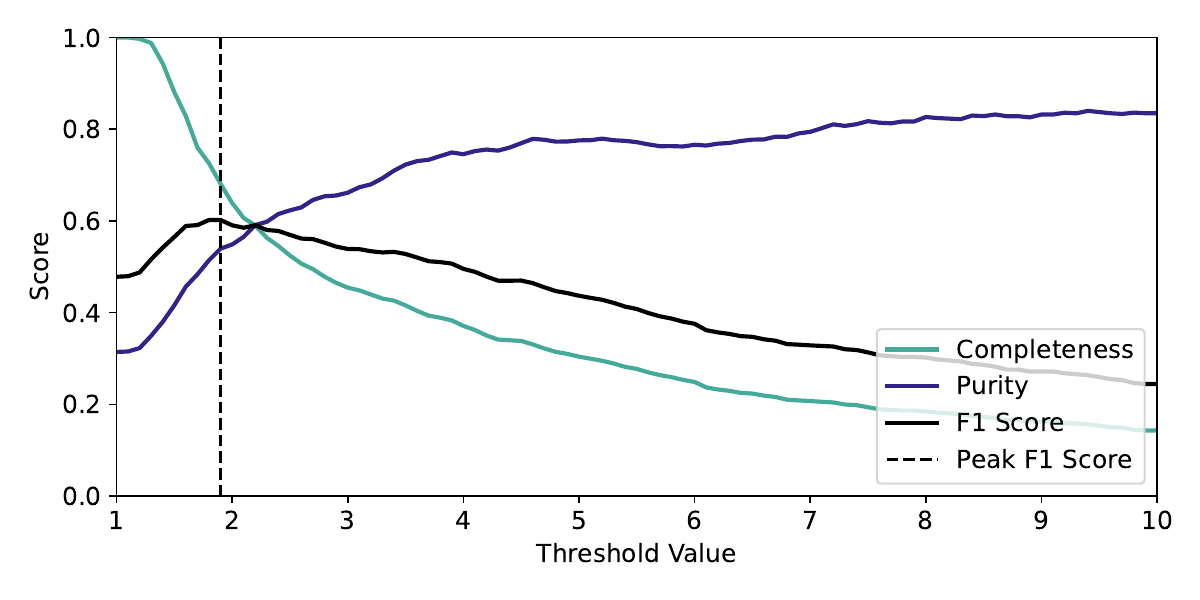}
    \caption{Plot of completeness, purity and F1 score against host contrast parameter threshold value. The F1 score peaks at a value of 0.60 at threshold value 1.9 (indicated by the dashed line), with corresponding completeness and purity values of 0.68 and 0.53, respectively.}
    \label{fig:r_hc_cpf1}
\end{figure}

This appendix contains all of the corresponding $r$ band plots for the tests carried out on the $g$ band photometry in Section \ref{Section2}. The results are broadly similar to those in the $g$ band. The $r$ band distribution of the log ratio of first detection flux to the pre-detection standard deviation is shown in Figure \ref{fig:r_std_hist}, and the completeness, purity and F1 scores produced by applying a range of thresholds are shown in Figure \ref{fig:r_std_cpf1}. 
The $r$ band distribution of the log ratio of first detection flux to the pre-detection mean flux is shown in Figure \ref{fig:r_mf_hist}, and the completeness, purity and F1 scores produced by applying a range of thresholds are shown in Figure \ref{fig:r_mf_cpf1}. 
The $r$ band distribution of the log ratio of first detection flux to the pre-detection root mean square is shown in Figure \ref{fig:r_rms_hist}, and the completeness, purity and F1 scores produced by applying a range of thresholds are shown in Figure \ref{fig:r_rms_cpf1}. 
The $r$ band distribution of the log of the host contrast parameter defined in \citet{Hung2018} is shown in Figure \ref{fig:r_host_contrast_hist}, and the completeness, purity and F1 scores produced by applying a range of thresholds are shown in Figure \ref{fig:r_hc_cpf1}. 

\section{Light Curve History Dependence of Other Parameters}
\label{APP_lc_history}
Similarly to Section \ref{Section3.2}, we evaluated the impact of light curve history availability on the effectiveness of the other photometric variability parameters listed in Section \ref{Section2}. We do not assess the impact on the host contrast parameter as it is not effected by the quantity of light curve history. For each parameter, we calculate the ROC curves for the full light curve history, 180 days of light curve history and 30 days of light curve history.

\begin{figure}
	\includegraphics[width=\columnwidth]{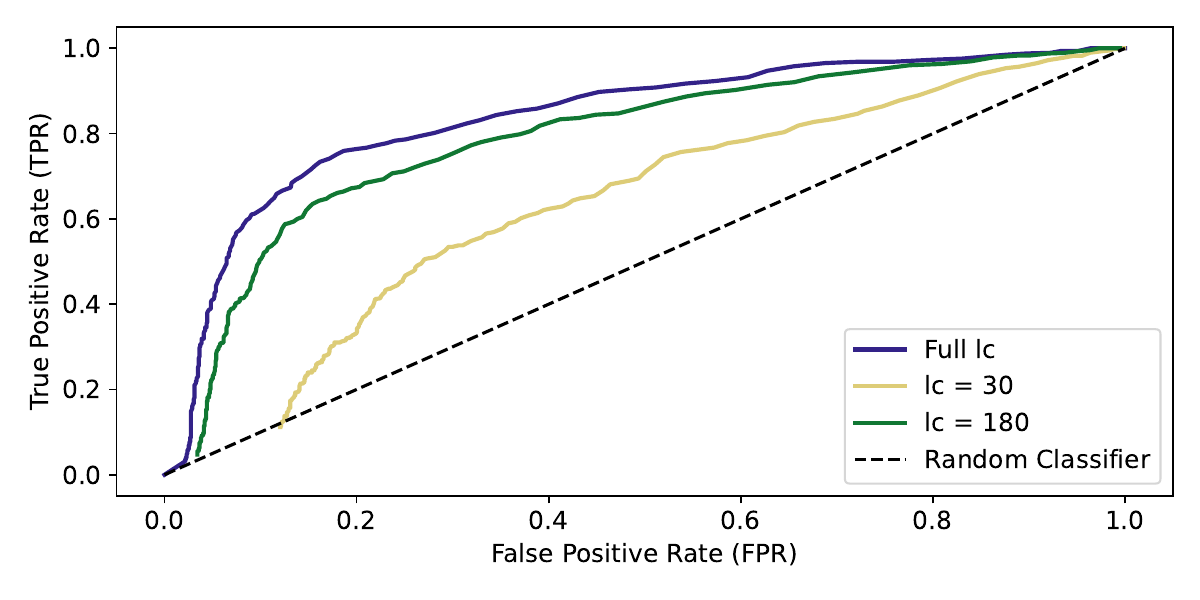}
    \caption{ROC curves for the cut using the ratio of detection flux to pre-detection standard deviation calculated with the following amounts of ZTF light curve history: full light curve, 30\,d pre-detection and 180\,d pre-detection. The larger the area under the curve, the more effective the parameter cut is at distinguishing between non-AGN transients and AGN. The dashed black line indicates the expected performance of a purely random classifier. As is expected, the cut is most effective when the most history is available. There is a distinct drop in effectiveness for this parameter in isolation when the amount of light curve history is limited to 30\,d.}
    \label{fig:sd_ROC_time}
\end{figure}

The impact of light curve history on the effectiveness of using the ratio of detection flux to pre-detection standard deviation to separate AGN from non-AGN is shown in Figure \ref{fig:sd_ROC_time}. As is expected, the cut is most effective when the most history is available. There is a distinct drop in effectiveness for this parameter in isolation when the amount of light curve history is limited to 30\,d.

\begin{figure}
	\includegraphics[width=\columnwidth]{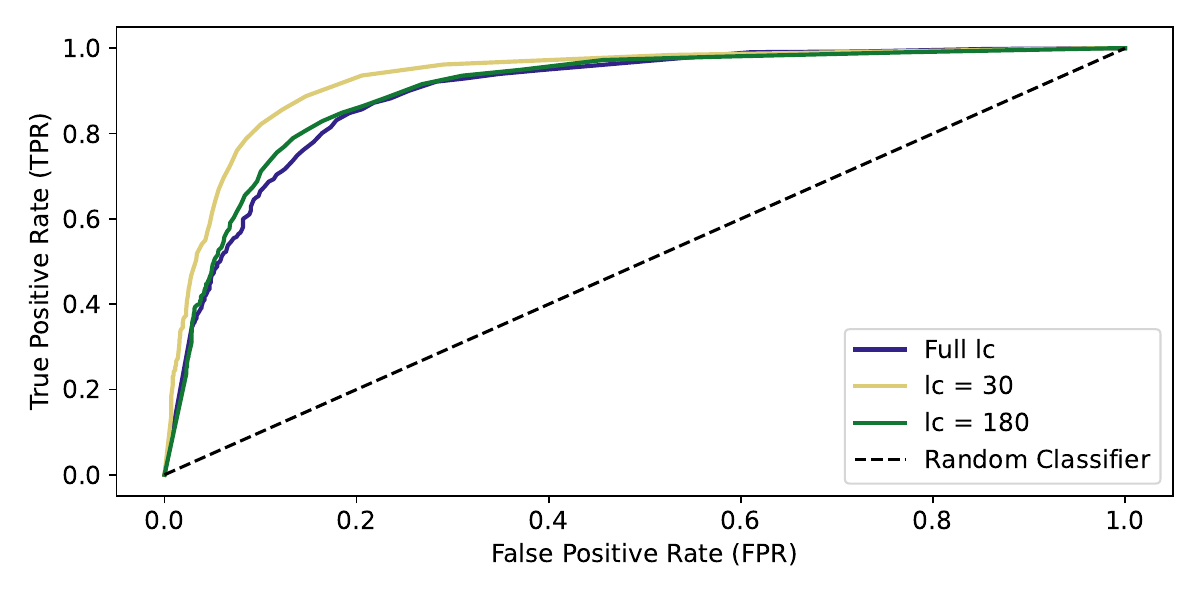}
    \caption{ROC curves for the cut using the ratio of detection flux to pre-detection mean flux calculated with the following amounts of ZTF light curve history: full light curve, 30\,d pre-detection and 180\,d pre-detection. The dashed black line indicates the expected performance of a purely random classifier. When applied in isolation, this parameter cut appears to be more effective when less light curve history is used. This is possibly due to AGN variability being more pronounced within the mean when there are less overall data points.}
    \label{fig:mf_ROC_time}
\end{figure}

The impact of light curve history on the effectiveness of using the ratio of detection flux to pre-detection mean flux to separate AGN from non-AGN is shown in Figure \ref{fig:mf_ROC_time}. When applied in isolation, this parameter cut appears to be more effective when less light curve history is used. This is possibly due to AGN variability being more pronounced within the mean when there are less overall data points.

\begin{figure}
	\includegraphics[width=\columnwidth]{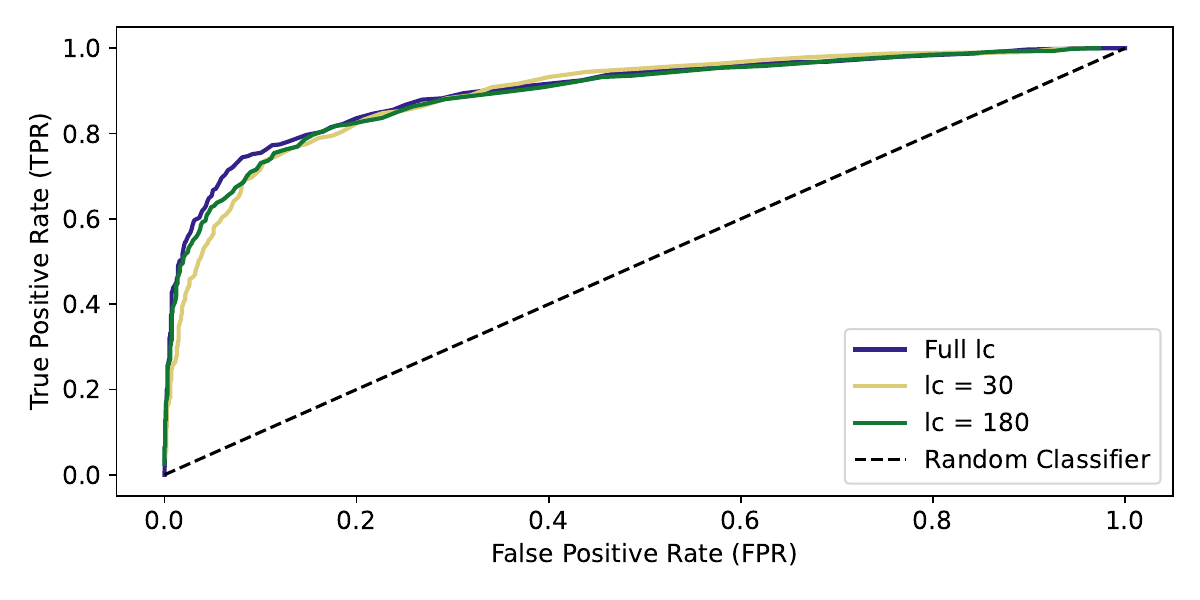}
    \caption{ROC curves for the cut using the ratio of detection flux to pre-detection root mean square calculated with the following amounts of ZTF light curve history: full light curve, 30\,d pre-detection and 180\,d pre-detection. The dashed black line indicates the expected performance of a purely random classifier. As is expected, the cut is most effective when the most history is available. However, the effect of limiting the quantity of light curve history appears to be more minimal for this parameter compared to others, as it still remains very effective with limited history.}
    \label{fig:rms_ROC_time}
\end{figure}

The impact of light curve history on the effectiveness of using the ratio of detection flux to pre-detection root mean square to separate AGN from non-AGN is shown in Figure \ref{fig:rms_ROC_time}. As is expected, the cut is most effective when the most history is available. However, the effect of limiting the quantity of light curve history appears to be more minimal for this parameter compared to others, as it still remains very effective with limited history.

\section{LSST Filter Choice Experimentation}
\label{APP_filter_choice}

We evaluate the impact of using different combinations of LSST filters ($u$, $g$, $r$, $i$, $z$ \& $y$) for the two-dimensional cut to distinguish non-AGN transients from standard AGN variability. The different filters each have different depths and frequency of inclusion within the LSST observing cadence. Therefore, it is reasonable to expect that some bands will be of more utility in determining the variability of the source than others. For example, greater error in one of the shallower bands could result in artificial inflation of the variability values, resulting in the rejection of an object that should otherwise pass.

We test five different combinations of filters ($gri$, $ugri$, $griz$, $ugriz$ \& $ugrizy$) on the MALLORN data set. The results are shown in an ROC curve in Figure \ref{APP_filter_choice}. $gri$ is the most effective combination, producing the largest area under the curve, likely due to the depth of these bands ($g_{lim} = 24.5$, $r_{lim} = 24.0$, $i_{lim} = 23.4$) and their more frequent inclusion in the LSST cadence, as for a given night of observations there is at least one pair of $g+r$, $g+i$ or $r+i$. $ugri$ and $griz$ are slightly less effective than just $gri$ but are still largely successful at distinguishing non-AGN transients from standard AGN variability. $ugriz$ is slightly less effective again, whilst $ugrizy$ is the least effective of the combinations, likely due to the shallow depth of the $y$ band ($y_{lim} = 22.0$).

\begin{figure}
	\includegraphics[width=\columnwidth]{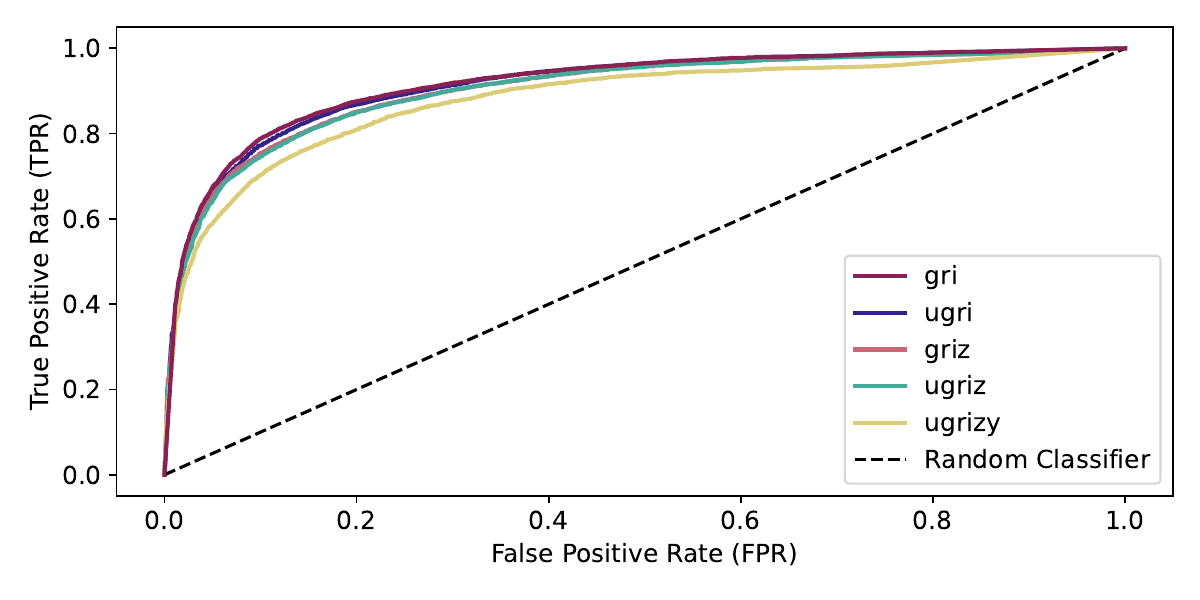}
    \caption{ROC curves comparing the effectiveness of using different filter combinations to apply the multi-dimensional cut to the MALLORN data set to distinguish non-AGN transients from AGN. Using just $g$, $r$ \& $i$ is the most effective combination, producing the largest area under the curve.}
    \label{fig:filter_choice}
\end{figure}


\bsp	
\label{lastpage}
\end{document}